\documentclass[aps,prl,twocolumn,showpacs,superscriptaddress,floatfix, usepdftitle=true]{revtex4-2}
\usepackage[utf8]{inputenc}
\usepackage{ucs}
\usepackage{natbib}
\usepackage{graphicx}
\usepackage{bm}
\usepackage{amssymb}
\usepackage{amsmath}
\usepackage{color}
\usepackage{upgreek}
\usepackage{braket}
\usepackage{soul}
\usepackage{hyperref}
\usepackage{xcolor}
\hypersetup{
    colorlinks=true,
    linkcolor=blue,
    filecolor=magenta,      
    urlcolor=cyan,
    pdftitle={Liebman et al.},
    pdfpagemode=FullScreen}

\newcommand{\cm}{~\text{cm}$^{-1}$~}

\begin{document}
 
\title{Tuning Charge Order in Mott insulators $\kappa$-(BEDT-TTF)$_2$Hg(SCN)$_2$X (X= Br, Cl) via Uniaxial Strain}

\author{Jesse Liebman}
\affiliation{Department of Physics and Astronomy, The Johns Hopkins University, Baltimore, Maryland 21218, USA}

\author{Svetlana Torunova}
\affiliation{Institute of Problems of Chemical Physics, Chernogolovka, Russia}

\author{John A. Schlueter}
\affiliation{Division of Materials Research, National Science Foundation, Alexandria, Virginia 22314, USA}

\author{Elena Zhilyaeva}
\affiliation{Institute of Problems of Chemical Physics, Chernogolovka, Russia}

\author{Natalia Drichko}
\affiliation{Department of Physics and Astronomy, The Johns Hopkins University, Baltimore, Maryland 21218, USA}
\email{drichko@jhu.edu}

\date{\today}

\begin{abstract}

In condensed matter physics, experimental control over material properties reflects a deep understanding of the underlying physics. 
In recent years, meaningful progress has been made towards a description of the physics of correlated electron systems, but examples of control of these systems remain rare. 
In this work, we confirm a phase diagram theoretically proposed for organic Mott insulators. We use  $\kappa$-(BEDT-TTF)$_2$Hg(SCN)$_2$X (X=Br,Cl) (BEDT-TTF = bis(ethylenedithio)tetrathiafuvalene)  materials as experimental realization of the proposed model and demonstrate the ability to tune them both ways across a phase border between a Mott insulator with a uniformly distributed charge and a charge ordered state through the application of uniaxial strain. We induce charge order at 33~K in the quantum dipole liquid material $\kappa$-(BEDT-TTF)$_2$Hg(SCN)$_2$Br through the application of tensile strain of 0.4\% along the c-axis. We suppress charge order down to 10~K in $\kappa$-(BEDT-TTF)$_2$Hg(SCN)$_2$Cl by applying a tensile strain of 1.6\% along the b-axis. We use Raman scattering spectroscopy to probe the charge state through analysis of charge sensitive molecular vibrations
and a low frequency mode of collective dipole fluctuations close to the phase border.

\end{abstract}

\maketitle

{\it Introduction---} Control of unconventional collective behavior in correlated electron systems requires the tuning of the strength or symmetry of the underlying interactions. Models describing the ground states of correlated materials provide a parametrization of a phase diagram and suggest a pathway to transform the collective electronic ground state. Physical realization of such control calls for a material that exists close to a phase boundary which is defined by an experimentally tunable parameter. Materials which host competing interactions produce exotic phase diagrams, where an understanding of the hierarchy of interactions can be reached by tuning the ground state through the application of external stimuli. 

\begin{figure}
    \centering
    \includegraphics[width=\linewidth]{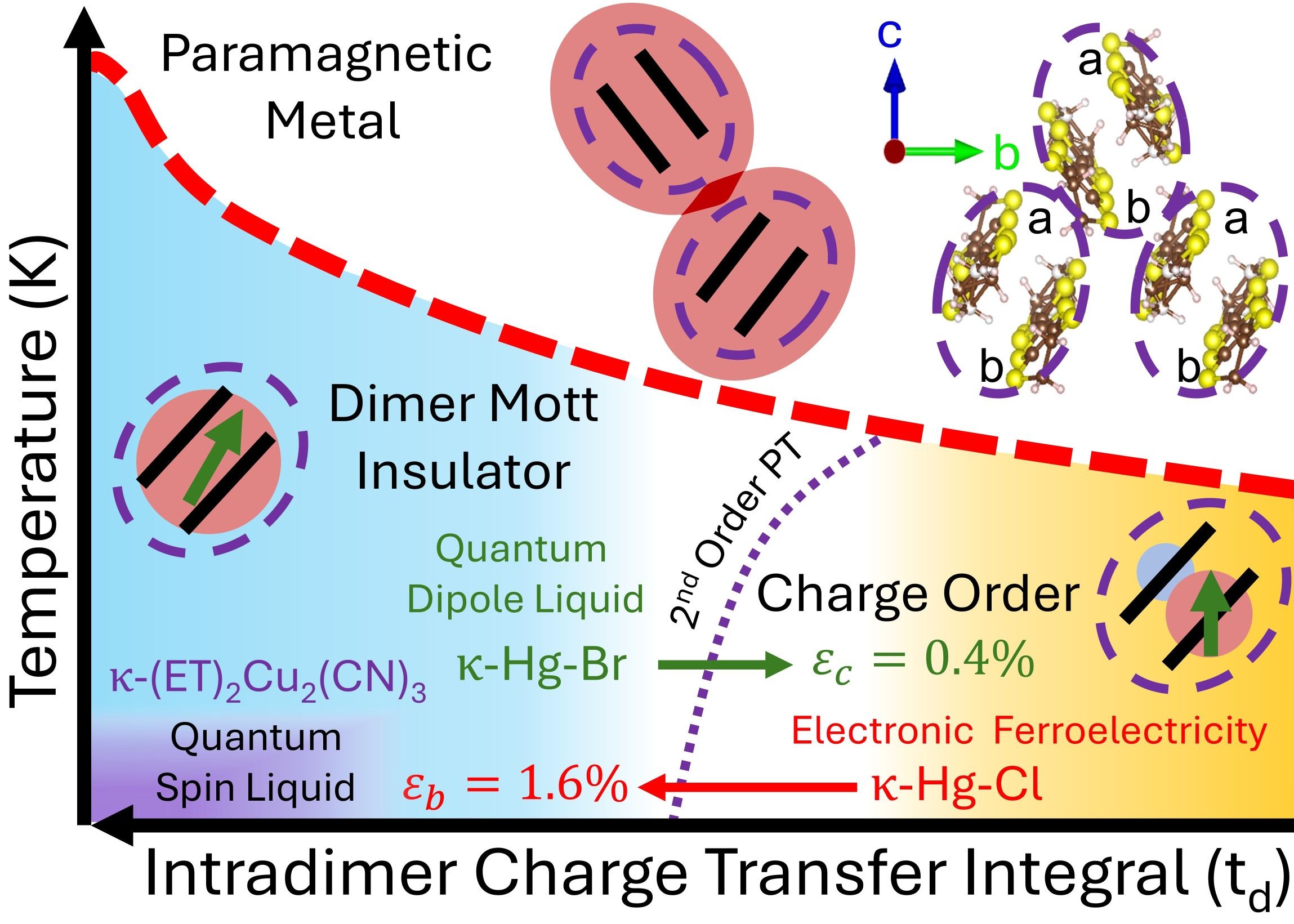}
    \caption{Phase diagram suggested for organic Mott insulators~\cite{Hotta2010,Naka2010} with experimental results for  $\kappa$-(BEDT-TTF)$_2$X, X=Cl, Br mapped on it. Large values of t$_d$, left: The hole is equally distributed within (BEDT-TTF)$_2^{1+}$ dimers, the material is in the dimer Mott insulator phase and can realize quantum spin liquid  and  quantum dipole liquid. Small values of t$_d$, right: charge order phase. The arrows, with corresponding vales of strain, show the strain tuning realized in this work between the dimer Mott insulator and charge order phases.}
    \label{Dimer}
\end{figure}
Inspired by independent experimental observations, a phase diagram with a tuning from a spin liquid candidate to a non-magnetic ferroelectric has been theoretically predicted  for the triangular lattice organic Mott insulators of the formula $\kappa$-(BEDT-TTF)$_2$X~\cite{Hotta2010,Naka2010,Hassan2018,Hassan2020} (see Fig.~\ref{Dimer}). 
These materials host a triangular lattice of dimers of BEDT-TTF molecules (referred to as ET)~\footnote{BEDT-TTF refers to (bisethylenedithio)tetrathiafuvalene} that carry charge $+1e$ and quantum spin $S$=1/2, as shown in Fig.~\ref{Dimer}. Insulating ground states in these materials are generally well described by the extended Hubbard model with strong electron-electron repulsion~\cite{Fukuyama1996, Hotta2010, Naka2010, Dayal2011}. Within this suggested phase diagram, the intradimer charge transfer integral $t_d$ parametrizes the phase transition between the $\frac{1}{2}$-filled dimer Mott insulator, where the hole (charge $+1e)$ is equally shared between ET molecules within dimer lattice sites, and the $\frac{1}{4}$-filled charge order insulator~\cite{Hotta2010,Naka2010}, where the hole is unequally distributed between ET molecules.  
Materials with large $t_d$~\cite{Kandpal2009} such as $\kappa$-(ET)$_2$Cu$_2$(CN)$_3$~\cite{Shimizu2003, Yamashita2008, Nakamura2014, Hassan2018, Hassan2018Crystals} and $\kappa$-(ET)$_2$Ag$_2$(CN)$_3$~\cite{Shimizu2016}, show experimental evidence of $\frac{1}{2}$-filled dimer Mott insulators with antiferromagnetic interactions in a triangular lattice of $S=1/2$ spins. The antiferromagnetic $\kappa$-(ET)$_2$X materials are discussed in the context of triangular lattice spin liquids driven by ring exchange~\cite{Misguich1999, Motrunich2005, Hotta2012, Holt2014} and electric dipole fluctuations coupled to spins~\cite{Hotta2010, Jacko2020}. 
On the other hand materials with smaller $t_d$, such as the  $\kappa$-(ET)$_2$Hg(SCN)$_2$X (X=Br, Cl)~\cite{Aldoshina1993, Drichko2014, Gati2018}, are expected to reside closer to the charge order transition.  
$\kappa$-(ET)$_2$Hg(SCN)$_2$Br (referred to as $\kappa$-Hg-Br) demonstrates collective dipole fluctuations~\cite{Hassan2018} proposed for a quantum dipole liquid close to the phase border~\cite{Hotta2010,Naka2010}, as well as exotic magnetic behavior arising from charge fluctuations~\cite{Yamashita2021, Urai2022}. 
$\kappa$-(ET)$_2$Hg(SCN)$_2$Cl (referred to as $\kappa$-Hg-Cl) is a charge order insulator below T$_{\text{CO}}$=30~K~\cite{Drichko2014}, a realization of electronic ferroelectricity~\cite{Naka2010,Hotta2010,Gati2018}, and  exhibits signatures of a spin singlet state in the charge ordered regime~\cite{Drichko2022}. 
The chemical tuning of charge order within the $\kappa$-(ET)$_2$Hg(SCN)$_2$X (X=Br, Cl) family motivates this study of controlling the charge order transition through the application of uniaxial strain. We seek to realize both the dimer Mott insulator and charge order insulator as the ground states of each material by tuning the underlying interactions, robustly confirming the proposed phase diagram in Fig.~\ref{Dimer}. 

\begin{figure*}
    \includegraphics[width=\linewidth]{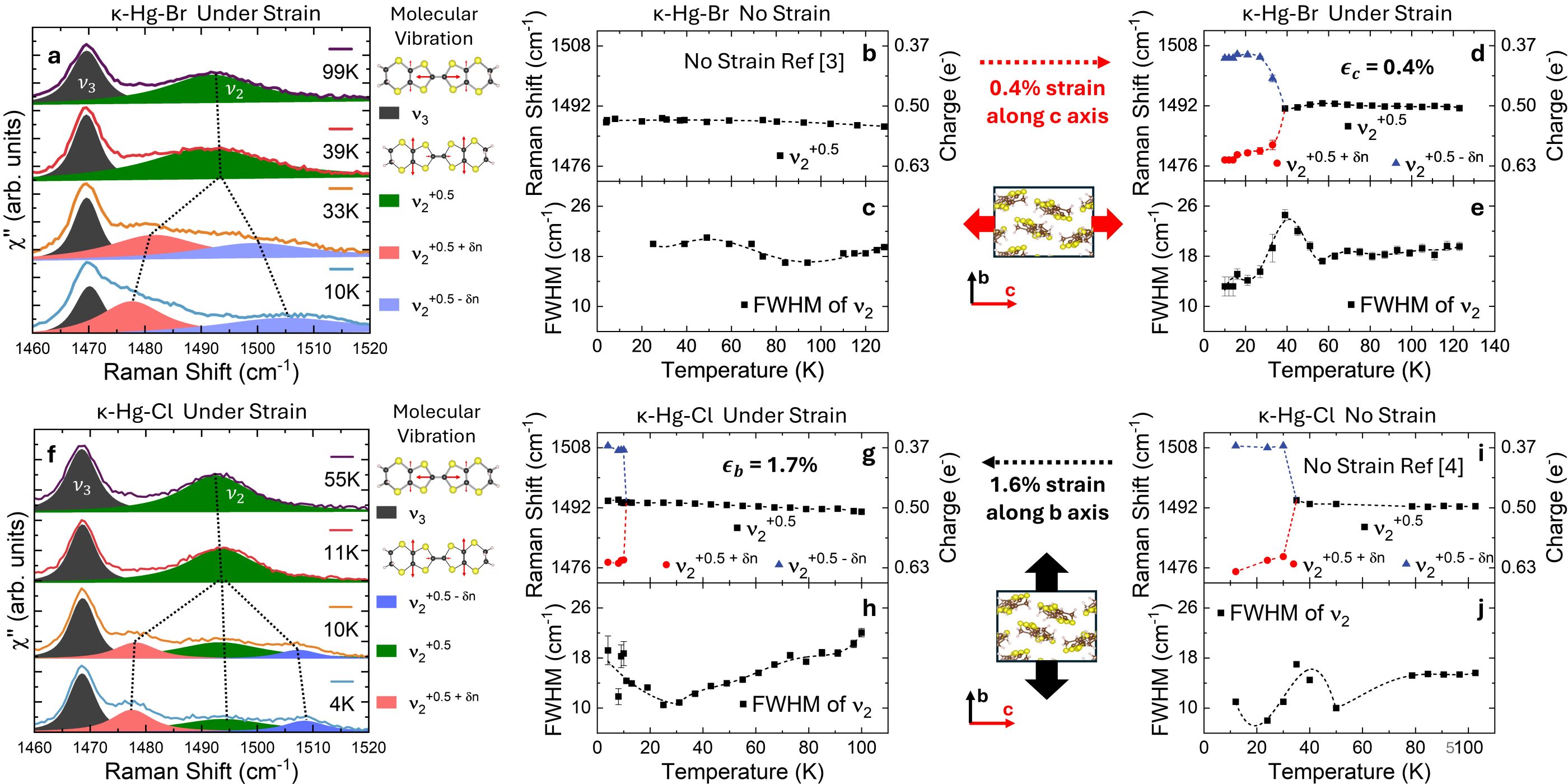}
    \caption{Effects of the application of strain as detected by the charge sensitive molecular vibrations of  $\kappa$-(BEDT-TTF)$_2$Hg(SCN)$_2$X (X=Br, Cl). For $\kappa$-(BEDT-TTF)$_2$Hg(SCN)$_2$Br: {\bf (a)} Raman scattering spectra under 0.4\% tensile strain along the c-axis, the {\bf (b)} frequency and {\bf (c)} full-width-half-maximum (FWHM) of $\nu_2$ without strain from Ref.~\cite{Hassan2018}, and the {\bf (d)} frequency and {\bf (e)} full-width-half-maximum (FWHM) of $\nu_2$ under 0.4\% tensile strain along the c-axis. For $\kappa$-(BEDT-TTF)$_2$Hg(SCN)$_2$Cl: {\bf (f)} Raman scattering spectra of under 1.6\% tensile strain along the b-axis, the {\bf (g)} frequency and {\bf (h)} full-width-half-maximum (FWHM) of $\nu_2$ under 1.6\% tensile strain along the b-axis, and the {\bf (i)} frequency and {\bf (j)} full-width-half-maximum (FWHM) of $\nu_2$ without strain from Ref.~\cite{Hassan2020}.}
    \label{Mol-Vib}
\end{figure*}

In recent years, strain tuning has been established as a useful method for control of many body quantum systems. 
To explore the variety of materials and physics which can be studied under strain, many recent advancements have been made in the methods and devices for strain application~\cite{Blundo2021, Kim2023}.
Band structure~\cite{Schmidt2016, Shin2016} and band topology~\cite{Zhao2020, Lin2021} have been successfully tuned under strain. Moreover, magnetic ground states~\cite{Cenker2022} have been reversibly tuned under uniaxial strain. The competition between nematicity and superconductivity in Fe-based superconductors has been studied under strain~\cite{Bohmer2017, Hristov2019, Malinowski2020, Ghini2021}. Recent thermodynamic measurements on $\kappa$-(ET)$_2$X materials under strain demonstrate the ability to control the temperatures of phase transitions of these materials~\cite{Lieberich2025}. While these studies have established strain as a key tuning parameter for electronic phase transitions, the ability to transform the electronic ground state across a phase boundary in both directions remains to be realized.

In this letter, we demonstrate  experimental evidence of tuning the ground states of $\kappa$-Hg-X via uniaxial strain across the phase border in two directions, both {\it in} and {\it out} of the charge ordered state.
We utilize the established method of characterization of the charge state of BEDT-TTF based materials through the analysis of charge sensitive molecular vibrations to identify the charge state at each temperature.
We observe in $\kappa$-Hg-Br softening of collective dipole fluctuations through the dimer Mott insulator to charge order insulator transition, suggesting collective behavior drives the transition.
Our results provide the first robust experimental evidence of the theoretical phase diagram in Fig.~\ref{Dimer}, demonstrate the tuning of ground states of strongly correlated electron systems with a novel mechanism, 
and serve as a foundation for tuning the collective behavior of these multiferroic materials.

{\it Experimental Procedure---} Crystals of ET-based materials are soft and fragile, which is reflected in large rates of thermal expansion~\cite{SI}. Therefore, we utilize a simple setup to apply thermal strain, using the large mismatch in the thermal expansion of $\kappa$-Hg-X when compared to metals such as Cu~\cite{Rubin1954, Drichko2014, SI}.
We expect the temperature-dependent strain to be constant below 100~K. Raman scattering spectroscopy is used to measure molecular vibrations, lattice phonons, and collective dipole fluctuations. The detailed experimental methods are provided in the Supplementary Material~\cite{SI}. 

We characterize the charge state of $\kappa$-Hg-X as a function of temperature and strain following analysis of high frequency molecular vibrations. We also measure the low frequency spectral range, where collective modes of charge fluctuations can be observed.

{\it Detecting Charge Order from Molecular vibrations---}
Charge order in the $\kappa$-(ET)$_2$X salts corresponds to an unequal distribution of charge between the individual ET molecules within (ET)$_2^{1+}$ dimers. As shown in Fig.~\ref{Dimer}, when the distribution of charge is equal ($n_a = n_b = 0.5$) the electronic system is in the dimer Mott phase, and unequal distribution of charge ($n_a\neq n_b$) corresponds to the charge order phase. To probe the charge state, we use the established method of characterization through the analysis of charge sensitive molecular vibrations~\cite{Yamamoto2005, Yakushi2012_Crystals}. The frequency of the Raman active molecular vibration $\nu_2$ depends linearly on the site charge $n$ as follows~\cite{Yamamoto2005}:
\begin{equation}
    \omega(n) = 1447\text{\cm} + 120 \text{\cm}(1-n)
    \end{equation}
At the charge order transition, the spectral profile of the $\nu_2$ vibration splits into two distinct lines with a frequency difference of $\Delta \omega$ corresponding to a  charge separation of
\begin{equation}
    \Delta n = |n_a - n_b| = \frac{\Delta \omega }{120 \text{\cm}}
\end{equation}
Due to such strong dependence on charge, fluctuations between the two differently charged molecules can broaden the vibrational line considerably. As a consequence, the linewidth of $\nu_2$ can be used as a tool to probe charge fluctuations near the charge order transition~\cite{Yakushi2012_Crystals,Hassan2018}.  
Previous studies of $\kappa$-Hg-Br at ambient pressure~\cite{Hassan2018} show a single spectral profile of $\nu_2$ (Fig.~\ref{Mol-Vib}{\bf (b)}) which broadens below 40~K (Fig.~\ref{Mol-Vib}{\bf (e)}), consistent with charge fluctuations in the dimer Mott phase, without reaching charge order on cooling.


We show the temperature dependent Raman scattering spectra of $\kappa$-Hg-Br under a tensile strain of 0.4\% along the $c$ axis in Fig.~\ref{Mol-Vib}{\bf(a)}. 
Under strain, the linewidth of $\nu_2$ increases below 60~K, suggesting the presence of charge fluctuations. 
The spectral profile of $\nu_2$ splits below T$_{\text{CO}}$=33~K, indicating the charge order transition.  Below the charge order transition temperature,  two distinct, narrow spectral profiles of $\nu_2$ are observed~\cite{Hassan2018, Hassan2020}, consistent with static charge order~\cite{Liebman2024}. From the frequencies of $\nu_2$ (see Fig.~\ref{Mol-Vib}{\bf(d)}),  we find that the charge separation is 0.24$e$, slightly smaller than the charge separation of 0.26$e$ found for $\kappa$-Hg-Cl in the charge order phase~\cite{Drichko2014, Hassan2020}.
See the Supplemental Material~\cite{SI} for a detailed description of the analysis of charge sensitive molecular vibrations.


Previous studies for $\kappa$-Hg-Cl at ambient pressure~\cite{Hassan2020} show a splitting of $\nu_2$ at 30~K, denoting the charge order transition (Fig.~\ref{Mol-Vib}{\bf (i)}). The linewidth increases above the transition and then drops significantly for the two components of $\nu_2$  below the charge order transition (Fig.~\ref{Mol-Vib}{\bf (j)}), consistent with static charge order.
We show the temperature dependent Raman scattering spectra of $\kappa$-Hg-Cl under a tensile strain of 1.6\% along the $b$ axis in Fig.~\ref{Mol-Vib}{\bf(f)}.
We do not observe a splitting of $\nu_2$ down to 11~K, which evidences the suppression of charge order.
At T$_{\text{CO}}$= 10~K, where $\nu_2$ finally splits, we observe 3 distinct peaks, evidence of a mixed state of (ET)$^{1+}_2$ dimer lattice sites with and without charge order, which persist down to 4~K. 
The estimated charge separation within charge ordered dimers is 0.26$e$ (Fig.~\ref{Mol-Vib}{\bf(g)}), identical to the charge separation without strain~\cite{Drichko2014, Hassan2020}.
Under strain (Fig.~\ref{Mol-Vib}{\bf(h)}) the linewidth begins to increases at 25~K, above the transition; however, the linewidth does not significantly decrease below the charge order transition. This suggests  the presence of an inhomogeneous mixture of (ET)$_2^{1+}$ dimer lattice sites with and without charge order.

{\it Collective dipole fluctuations---} 
It was demonstrated experimentally in Ref.~\cite{Hassan2018} under ambient strain below 80~K in the insulating dimer Mott state of $\kappa$-Hg-Br~\cite{Ivek2017}  that charge fluctuations, detected through the analysis of molecular vibrations, also result in a collective mode. This mode was  observed at a frequency around 50\cm, consistent with the frequency deduced independently from vibrational spectroscopy, and is in agreement with dipole liquid behavior~\cite{Yao2018, Hassan2018}. Indeed, theory predicts~\cite{Hotta2010, Naka2010} a collective excitation of such charge fluctuations, which behaves as a soft mode upon approaching the charge order transition, and appears at a finite frequency in the ordered state~\cite{Naka2013}. Unlike lattice-related soft modes of displacement type ferroelectrics, this mode does not correspond to lattice vibrations, but rather to collective fluctuations of charge within (ET)$_2^{1+}$ dimer lattice sites. 

\begin{figure}[t]
    \centering
    \includegraphics[width=\linewidth]{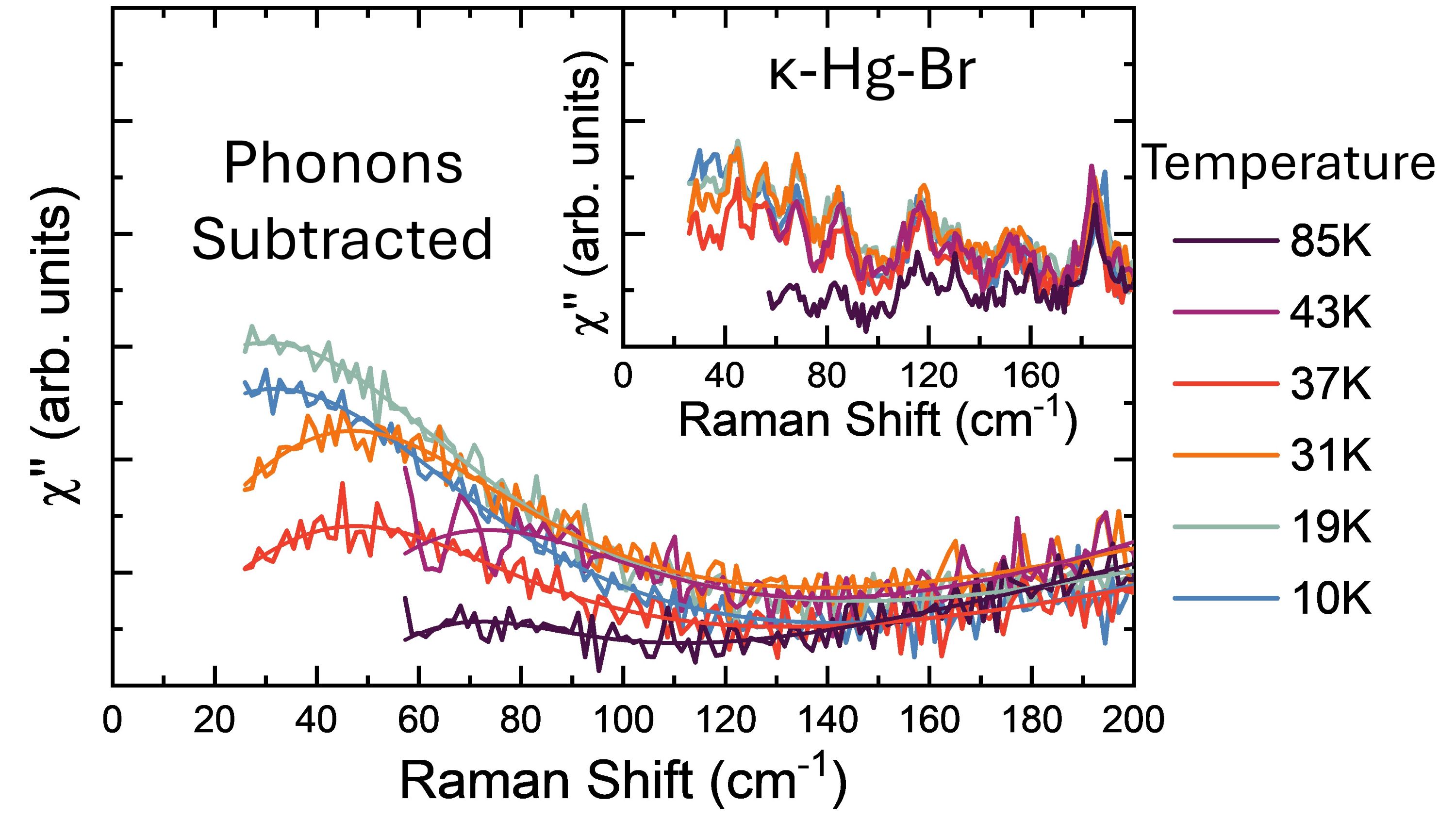}
    \caption{Low frequency Raman scattering spectra of $\kappa$-(BEDT-TTF)$_2$Hg(SCN)$_2$Br in the (b,b) polarization, with phonons subtracted. For each spectrum, the corresponding fitted Lorentzian profile is plotted. Inset: Original spectra including phonons.}
    \label{DF}
\end{figure}

We examine the Raman spectra of $\kappa$-Hg-Br under strain in the low frequency region below 200\cm. We plot the temperature dependent Raman spectra of $\kappa$-Hg-Br in the (b,b) polarization with phonons subtracted in Fig.~\ref{DF} (for the original spectra see inset). At temperatures below 80~K we observe a broad mode in this range, softening through the strain-induced charge order transition.
We fit this broad feature to a Lorentzian oscillator and plot the temperature dependence of the obtained frequency in Fig.~\ref{DF-Pars}.
For a complete description of the fitting procedure and parameter estimation, see the Supplemental Material~\cite{SI}.
At 37~K the center of the broad mode is close to 50\cm. Below 37~K down to 20~K, the spectral weight shifts to lower frequencies, down to around 30\cm. 
We report the integrated intensity of the collective mode at each temperature, $\text{I}(\text{T})$, which is a fit parameter of the Lorentzian profile.
As shown in the inset of Fig.~\ref{DF-Pars}, there is an increase in the integrated intensity of the dipole fluctuations mode on cooling.
The shift of spectral weight to lower frequencies is consistent with the behavior of a soft mode of collective dipole fluctuations through the charge order phase transition. 

In the absence of strain, $\kappa$-Hg-Cl undergoes a metal to charge order insulator transition at T$_{\text{CO}}$=30~K~\cite{Drichko2014}, where the collective excitation is not observed~\cite{Hassan2018Crystals}. See the supplementary material (Section IV and Fig.~5)~\cite{SI} for the corresponding spectra. Under uniaxial strain, while the presence of the elastic line precludes us from going to low frequencies, the changes in the observed range are negligible.
The absence of a collective mode is consistent with the mixed state deduced from the analysis of molecular vibrations.
\begin{figure}
    \centering
    \includegraphics[width=\linewidth]{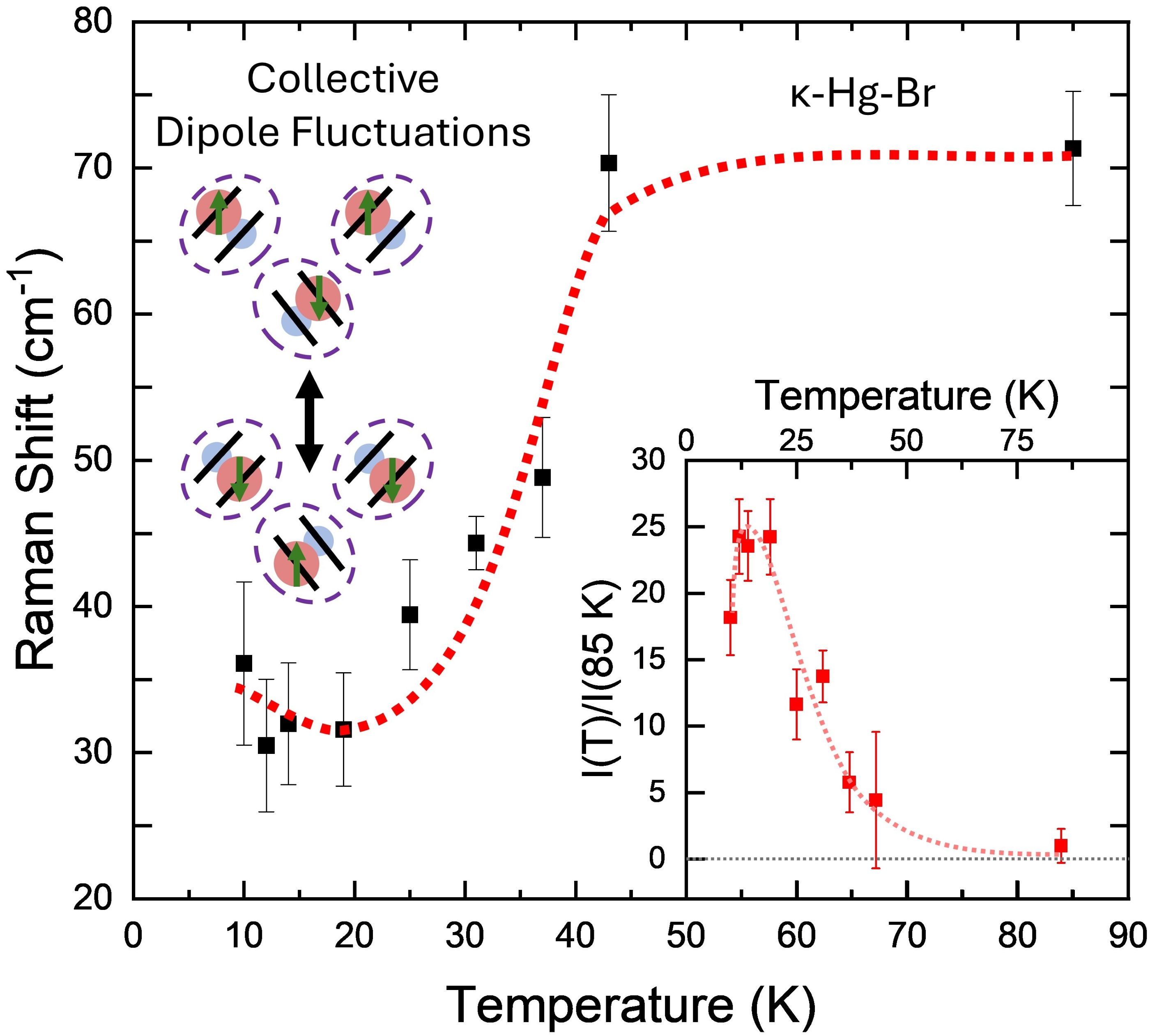}
    \caption{Frequency of collective dipole fluctuations in $\kappa$-(BEDT-TTF)$_2$Hg(SCN)$_2$Br, obtained from the fit of the low-frequency Raman scattering spectra presented in Fig.~\ref{DF}. Inset: Temperature dependence of the integrated intensity of the collective mode. Intensities are normalized to the intensity at 85~K. Red dashed lines are guides to the eye.}
    \label{DF-Pars}
\end{figure}

{\it Discussion---}
Below we use calculations on the extended Hubbard model to understand the effect of strain on the ground state of the studied materials. Correlated electron physics phenomena in $\kappa$-(ET)$_2$X are best understood through the extended Hubbard model~\cite{Fukuyama1996,Hotta2010,Naka2010,Dayal2011}.
This model includes intradimer interactions as well as interdimer interactions. The intradimer interactions include intramolecular electron-electron repulsion $U$, the intradimer charge transfer integral $t_d$, and intradimer electron-electron repulsion $V_d$, which are renormalized to $U_{\text{eff}}$. When the intradimer charge transfer integral $t_d$ is large enough, the correlated electron system becomes $\frac{1}{2}$-filled in the dimer Mott phase.
The extended Hubbard model is then written as follows:
\begin{equation}\label{Eq. 8}
    \begin{split}
H_1 = \ & \ U_{\text{eff}}\sum_{\text{dimers}\atop i}  n_{i\uparrow}n_{i\downarrow} + \sum_{\text{dimers} \atop i\neq j}V^{ij}(n_{i\uparrow} + n_{i\downarrow})(n_{j\uparrow} + n_{j\downarrow})\\
& - \sum_{\text{dimers} \atop i\neq j}\sum_{\text{Spin }\sigma}t^{ij}(c^{\dagger}_{i\sigma}c_{j\sigma} + \text{h.c.})
\end{split}
\end{equation}
where long range electron-electron repulsion ($V^{ij}$) and charge transfer integrals ($t^{ij}$) can are expressed between each pair of dimer sites. $V$ and $V'$ correspond to electron repulsion between nearest neighboring and second nearest neighboring (ET)$_2^{1+}$ dimer sites, respectively. Likewise, $t$ and $t'$ correspond to charge transfer between nearest neighboring and second nearest neighboring (ET)$_2^{1+}$ dimer sites. The nearest neighbor interactions ($t, V$) are stronger than the second-nearest neighbor interactions ($t', V'$).

\begin{figure}[h]
    \centering
    \includegraphics[width=\linewidth]{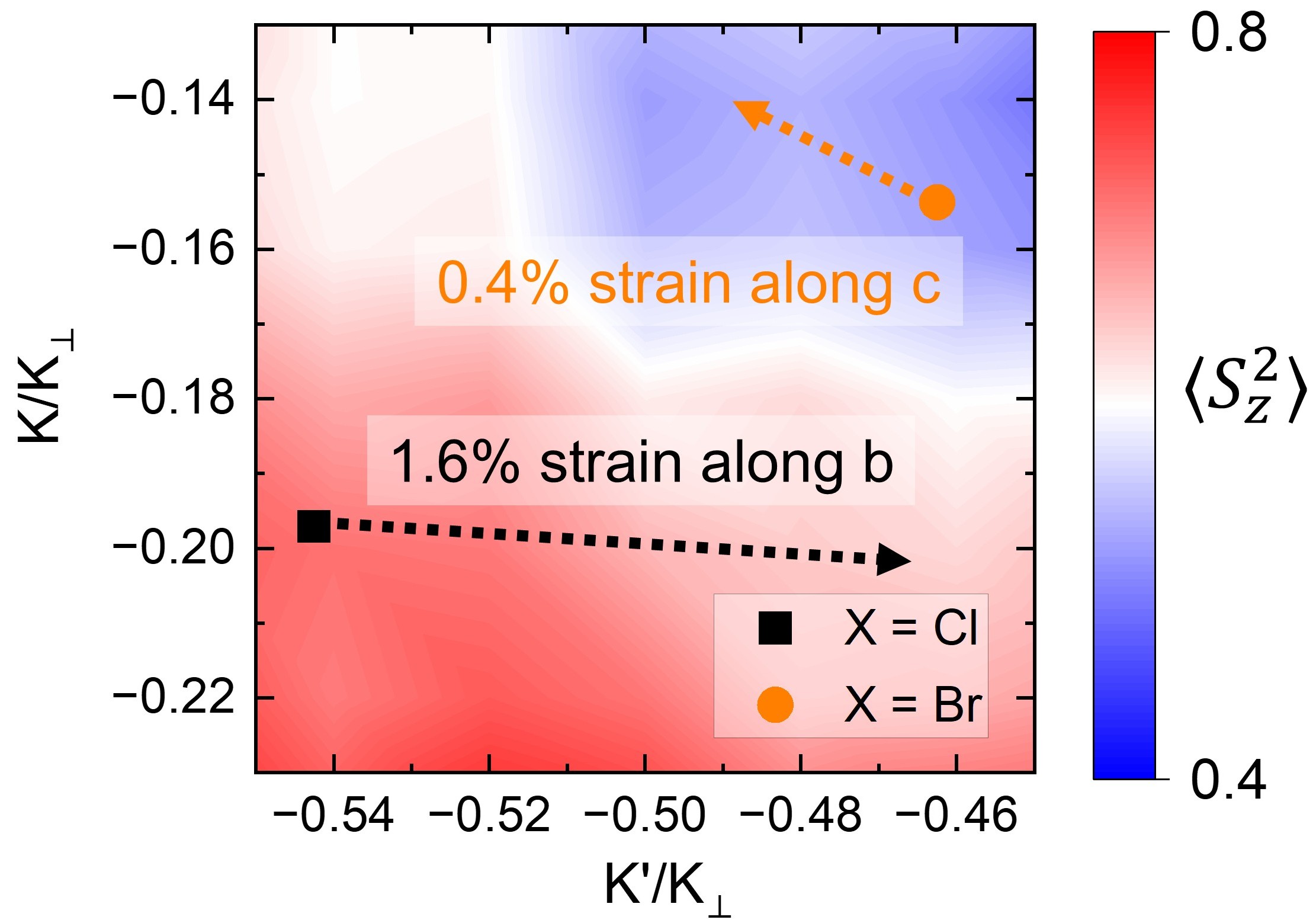}
    \caption{Phase Diagram of TFIM parameters from Ref.~\cite{Jacko2020}, with $\langle S_z^2 \rangle $ estimated via Monte Carlo Simulation.}
    \label{TFIM_PD}
\end{figure}

Theoretical predictions using the extended Hubbard and Kugel-Khomskii Models~\cite{Hotta2010, Naka2010, Naka2013, Jacko2020} parameterize the charge order transition in terms of competition between interdimer electron repulsion, $V$, and intradimer charge transfer, $t_d$. 
Following Ref.~\cite{Naka2010, Hotta2010, Jacko2020} the emergent electric dipoles on dimer sites can be mapped onto Ising-like dipoles represented by $S=1$ spins and subsequently modeled by the $S=1$ transverse field Ising model (TFIM):
\begin{equation}
    \hat{H} \ = \ K_\perp\sum_i\hat{S}^x_i + \frac{1}{2}\sum_{i\neq j}K^{ij}\hat{S}^z_i\hat{S}^z_j
    \label{TFIM-2}
\end{equation}
Where $K_\perp=2t_d$ is the effective transverse field and $K^{ij}$ ($K$ or $K'$) is the coupling between electric dipoles that are represented as $S=1$ spins. 
We perform a Monte Carlo simulation of the TFIM in order to evaluate the expectation value of $\langle \hat{S}_z^2 \rangle$, which corresponds to the spontaneous electric dipole moment presented in the phase diagram in Fig.~\ref{TFIM_PD}.
While our calculations do not quantitatively reproduce the experimentally observed values of charge separation in the studied materials, they qualitatively explain the observed trends in the data and identify the parameters which control the ground state. Future theoretical investigation, implementing higher order effects such as dipole-spin coupling~\cite{Hotta2010, Jacko2020}, could precisely and quantitatively explicate the strain tuning of charge order.

\begin{figure}[h]
    \centering
    \includegraphics[width=0.8\linewidth]{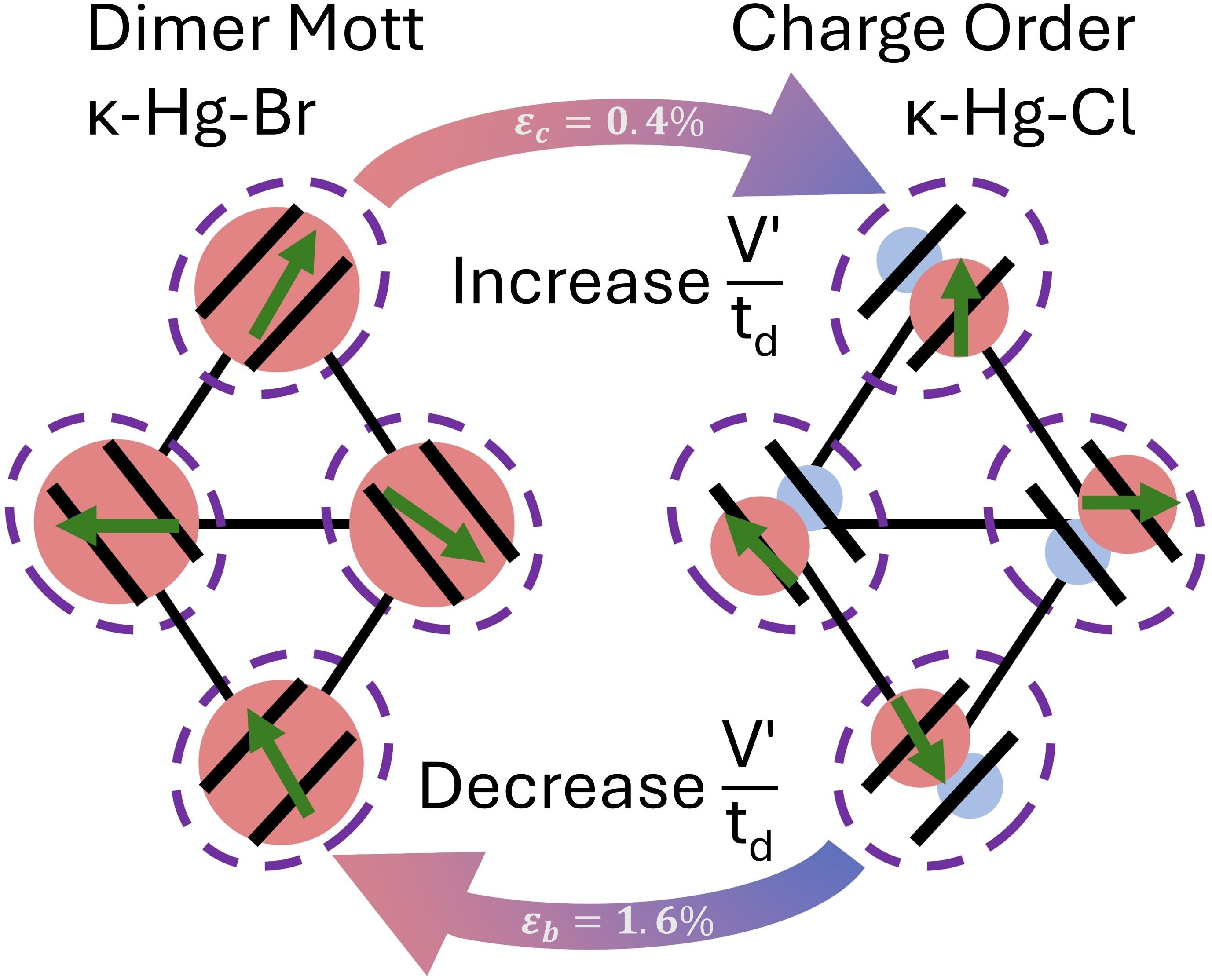}
    \caption{Strain tuning of charge order in $\kappa$-(BEDT-TTF)$_2$Hg(SCN)$_2$X (X=Br, Cl). The tuning of the ratio $V'/t_d$ is suggested as a pathway for control the charge order phase transition.}
    \label{Strain Tuning}
\end{figure}

Our analysis shows that under strain the second-nearest neighbor interaction $K'$ undergoes substantial tuning when compared to the nearest neighbor interaction $K$.
$K_\perp = 2t_d$ does not lie entirely along one axis, but it is expected that tensile strain along either crystallographic axis will slightly decrease $t_d$.
Application of tensile strain on $\kappa$-Hg-Br along $c$-axis leads to an increase of the  $|K'/K_\perp|$ ratio,  driving a larger expectation value of $\langle S_z^2 \rangle$, which means that charge order becomes more energetically favorable.
Application of tensile strain on $\kappa$-Hg-Cl along $b$-axis leads to  a decrease of the  $|K'/K_\perp|$ ratio, leading to a smaller expectation value of $\langle S_z^2 \rangle$, and indicating that charge order becomes less energetically favorable.
This analysis demonstrates  that for the $\kappa$-Hg-X materials under strain, the charge order transition is driven largely by the competition between $K_\perp \propto t_d$ and $K'$, which is parametrized by interactions projected entirely onto the b-axis, namely $V'$. The  compressive strain along $b$-axis strengthens $K' \propto V'$ and tensile strain along $b$-axis weakens $K' \propto V'$. We therefore conclude that charge order in $\kappa$-Hg-X salts is tuned on the application of strain through the ratio of $V'/t_d$, as shown in Fig.~\ref{Strain Tuning}.

{\it Conclusions---}
We provide experimental evidence of tuning the triangular lattice Mott insulators $\kappa$-(BEDT-TTF)$_2$X materials in both directions across the charge order phase border by uniaxial strain, acting opposite to the ``chemical" strain (Fig.~\ref{Dimer}). 
Analysis of the frequency and linewidth of Raman-active charge sensitive molecular vibrations allows us to identify the charge state at each measured temperature and strain.
We successfully induced charge order in the quantum dipole liquid candidate $\kappa$-Hg-Br below 33~K by application of 0.4\% strain along $c$-axis, and successfully suppressed charge order in $\kappa$-Hg-Cl down to 10~K, below which a mixed state is observed, by application of 1.6\% uniaxial strain along $b$-axis.
The low frequency Raman response leads to the observation of a mode of collective charge fluctuations in $\kappa$-Hg-Br, while the mode is absent in $\kappa$-Hg-Cl, in agreement with the mixed charge state.
Our calculations, based on estimations of changes to the structural parameters with strain and using the transverse field Ising model to estimate charge order, establish a mechanism to control charge order via strain in all $\kappa$-(ET)$_2$X materials by tuning the ratio of $\frac{V'}{t_d}$ (Fig.~\ref{Strain Tuning}).

We establish the control of electronic ferroelectricity through modification of competing interactions of emergent electric dipoles. We successfully tune the charge order transition in $\kappa$-(BEDT-TTF)$_2$X materials with external strain,  and we demonstrate successful tuning in both directions across the phase transition; therefore, we provide the first robust evidence of the proposed phase diagram in Fig.~\ref{Dimer}.
We hope that our work will serve as a foundation for tuning multiferroic materials with an optimistic outlook for switching behavior of multiple degrees of freedom.
\begin{acknowledgements}
{\it Acknowledgments---} The authors thank S. Ghosh, B. Powell, and S. Winter for fruitful discussions. The work at JHU was supported by NSF Award No. DMR-2004074. J.A.S. acknowledges support from the Independent Research/Development program while serving at the NSF. The preparation of single crystals was supported by the Ministry of Science and Higher Education of the Russian Federation (Registration number 124013100858-3). 
\end{acknowledgements}

\appendix*
\section{Appendix}
{\it Extended Hubbard Model---}
Within (ET)$^{+1}_2$ dimer lattice sites, the Hubbard model is written in terms of the intradimer interactions.
\begin{equation}\label{Eq. EM1}
    \begin{split}
H_0 = \ & \ U\sum_{\text{molecules}\atop i}  n_{i\uparrow}n_{i\downarrow} + V_d\sum_{\text{molecules} \atop i\neq j}(n_{i\uparrow} + n_{i\downarrow})(n_{j\uparrow} + n_{j\downarrow})\\
& - t_d\sum_{\text{molecules} \atop i\neq j}\sum_{\text{Spin }\sigma}(c^{\dagger}_{i\sigma}c_{j\sigma} + \text{h.c.})
\end{split}
\end{equation}
In the dimer Mott phase, the on-site Hubbard $U$ is renormalized to $U_{\text{eff}}$ which corresponds to the energy difference of 2 holes on separate dimer lattice sites and 2 holes on a single dimer lattice site:
\begin{equation}\label{Eq. EM2}
    U_{\text{eff}} = 2t_d+V_d+\frac{U-V_d}{2}\Biggl( 1-\sqrt{1+\Bigl(\frac{4t_d}{U-V_d}\Bigr)^2}\Biggr)
\end{equation}
In this case, the Hubbard model is written as Eq.~\ref{Eq. 8}. In the dimer Mott phase, the monomer terms are used to approximate the effective hopping terms as
\begin{equation}\label{Eq. EM3}
    t = \frac{1}{2}(t_{ab}+t_{aa}) \ \ \text{ and } \ \ t'=\frac{1}{2}(t'_{ab}+t'_{aa})=\frac{t'_{ab}}{2}
\end{equation} and the effective Coulombic interaction terms as 
\begin{equation}\label{Eq. EM4}
    V=\frac{1}{2}(V_{ab}+V_{aa}) \ \ \text{ and } \ \ V'=\frac{1}{2}(V'_{ab}+V'_{aa})=\frac{V'_{ab}}{2}
\end{equation}
\begin{figure}[h]
    \centering
    \includegraphics[width=\linewidth]{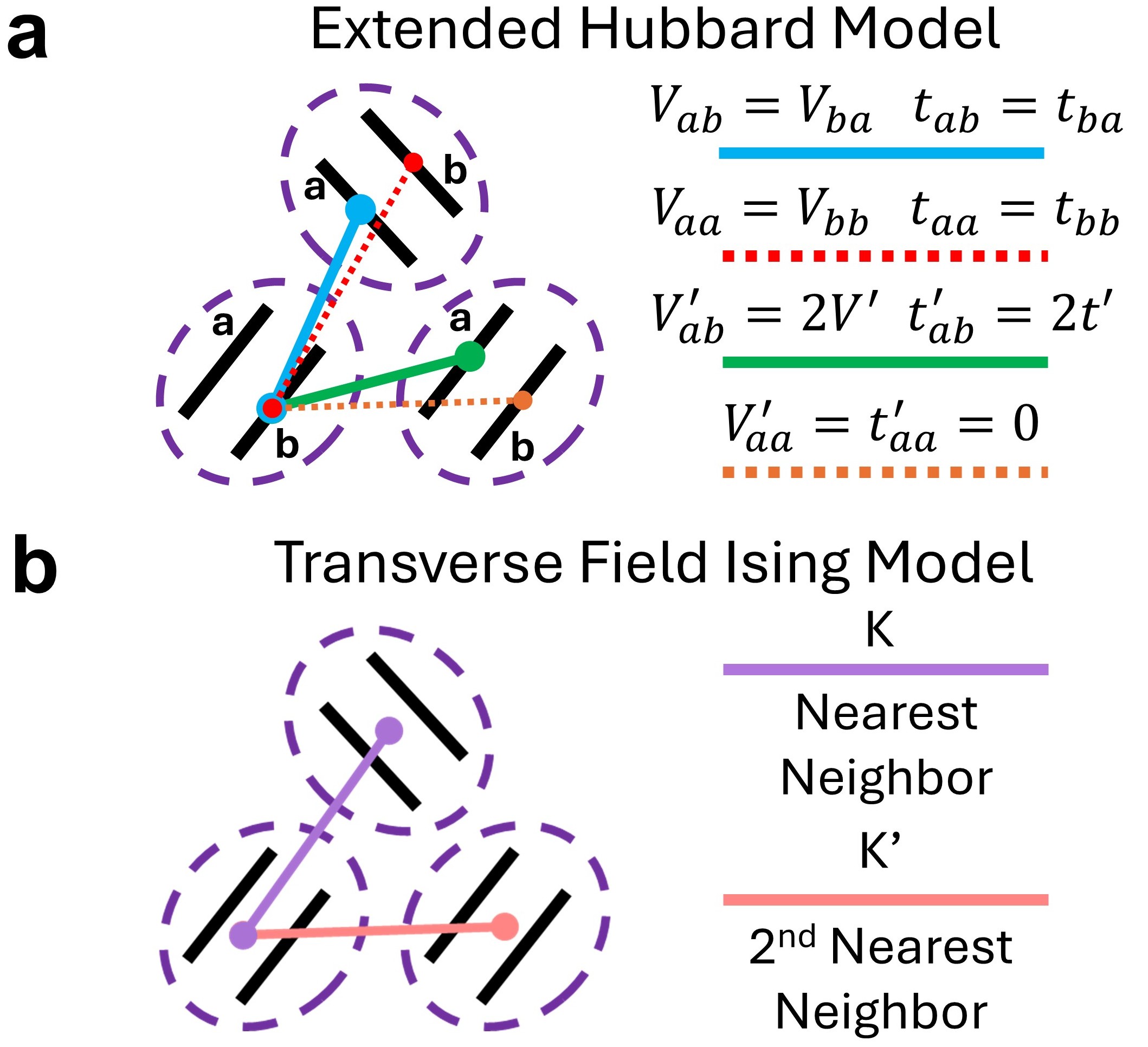}
    \caption{{\bf a} Monomer interdimer interactions of the extended Hubbard model. {\bf b} Interaction scheme of the transverse field Ising model (TFIM).}
    \label{Interaction}
\end{figure}

We show the monomer interactions of the extended Hubbard model in Fig.~\ref{Interaction}{\bf a}. In the charge order phase, $U_{\text{eff}}$ cannot be expressed by Eq.~\ref{Eq. EM2} and the approximations in Eqs.~\ref{Eq. EM3},~\ref{Eq. EM4} fail. In the charge order phase the monomer interactions are not simplified by symmetry; therefore, modeling the low energy physics of materials near the charge order phase transition is challenging.

{\it Transverse Field Ising Model---}
We visualize the mapping of the extended Hubbard model onto the transverse field Ising model in Fig.~\ref{Interaction}{\bf b}. 
The coupling between electric dipoles on nearest neighboring (ET)$_2^{1+}$ lattice sites is parametrized by $K$ and the coupling between second nearest neighbors is parametrized by $K'$ (see Fig.~\ref{Interaction}{\bf b}), analogous to electron repulsion between nearest ($V$) and second nearest ($V'$) neighboring (ET)$_2^{1+}$ dimer sites. The exact form of the $K^{ij}$ is written in the supplementary material~\cite{SI}, but the dependence of the $K^{ij}$ on the monomer interactions (see Fig.~\ref{Interaction}{\bf a}) is as follows:
\begin{equation}
    K^{ij} \propto V_{aa}^{ij}-V_{ab}^{ij} + \frac{(t_{aa}^{ij})^2-(t_{ab}^{ij})^2}{V_d}
\end{equation} 
It is important to note that $K^{ij}$ correspond to differences of the interdimer interactions between the $a$ and $b$ molecules within (ET)$_2^{1+}$ dimer sites, as shown in Fig.~\ref{Interaction}.
For $\kappa$-Hg-Br and $\kappa$-Hg-Cl the differences in these interactions between the $a$ and $b$ molecules are much larger for the second-nearest neighbor interactions ($t'$, $V'$) than the nearest neighbor interactions ($t$, $V$)~\cite{Koretsune2014, Jacko2020}.
Therefore, in the transverse field Ising model the interactions between second-nearest neighbors ($K'$) are expected to be twice as strong as nearest neighbors ($K$)~\cite{Jacko2020}, which is a region of parameter-space that lacks thorough investigation. We therefore turn to a detailed analysis of the tuning of the TFIM parameters under strain, using the monomer interactions from the extended Hubbard model. 
Using the crystallographic parameters of $\kappa$-Hg-X and their changes with the application of strain, we calculate and plot the tuning of the TFIM parameters in Fig.~\ref{Interaction}{\bf b} using simple approximations outlined in the Supplementary Material~\cite{SI}. 
For a detailed description of the analysis of the extended Hubbard model and TFIM under strain, see the Supplementary Material~\cite{SI}.

\section{Bibliography}
\bibliography{Arxiv.bbl}

\end{document}


\title{Supplemental Material: Tuning Charge Order in $\kappa$-(BED-TTF)$_2$Hg(SCN)$_2$X (X= Br, Cl) via Uniaxial Strain}

\author{Jesse Liebman}
\affiliation{Department of Physics and Astronomy, The Johns Hopkins University, Baltimore, Maryland 21218, USA}

\author{Svetlana Torunova}
\affiliation{Institute of Problems of Chemical Physics, Chernogolovka, Russia}

\author{John A. Schlueter}
\affiliation{Division of Materials Research, National Science Foundation, Alexandria, Virginia 22314, USA}

\author{Elena Zhilyaeva}
\affiliation{Institute of Problems of Chemical Physics, Chernogolovka, Russia}

\author{Natalia Drichko}
\affiliation{Department of Physics and Astronomy, The Johns Hopkins University, Baltimore, Maryland 21218, USA}
\email[]{drichko@jhu.edu}

\maketitle

\section{Raman scattering measurements}
Raman scattering spectra of single crystals of of $\kappa$-Hg-Br and $\kappa$-Hg-Cl in the frequency range from 12\cm to 2000\cm and temperature range from 300 K down to 6 K were measured using the Horiba Jobin-Yvon T64000 Triple Monochromator Spectrometer equipped with a liquid Nitrogen cooled CCD. Line of Ar$^+$-Kr$^+$ laser at 514.5nm was used for excitation.
Measurements were performed in a pseudo-Brewster angle geometry. We use a probe size of 100\um~$\times$ 50\um~with a power of 1.5~mW, for which we estimate laser heating of less than 1.5~K.
For low temperature measurement, samples were glued with GE varnish on a cold finger of a Janis ST500 Cryostat and cooled from 300 K to 5 K (3.5 K measured on the cold finger) using a cooling rate of 0.1~K/min. 

The crystals were oriented for measurements using by polarization dependent Raman scattering spectra.
Spectra presented in the manuscript were measured  in the (b,b) polarization for $\kappa$-Hg-Br and the (c,c) polarization for $\kappa$-Hg-Cl, allowing us to probe modes in the A$_g$ representation.  

Raman scattering spectra were normalized using Bose-Einstein statistics, following the fluctuation dissipation relation:
\begin{equation}
    \chi''(\omega,T)=\frac{S(\omega,T)}{1+n_{BE}(\omega,T)}
\end{equation}
where $\chi''(\omega,T)$ is the reported Raman scattering spectra for a given temperature $T$, $S(\omega,T)$ is the raw spectra measured at temperature $T$, and $n_{BE}(\omega,T)$ is the Bose-Einstein occupation for an excitation with energy $\hbar\omega$ at temperature $T$. For low frequency Raman scattering spectra, photoluminescence (PL) is subtracted from the data. The PL subtraction is described in section ``{\bf V. Treatment of Spectra}." 


\section{Application of strain}
\begin{figure*}[h]
    \centering
    \includegraphics[width=0.95\textwidth]{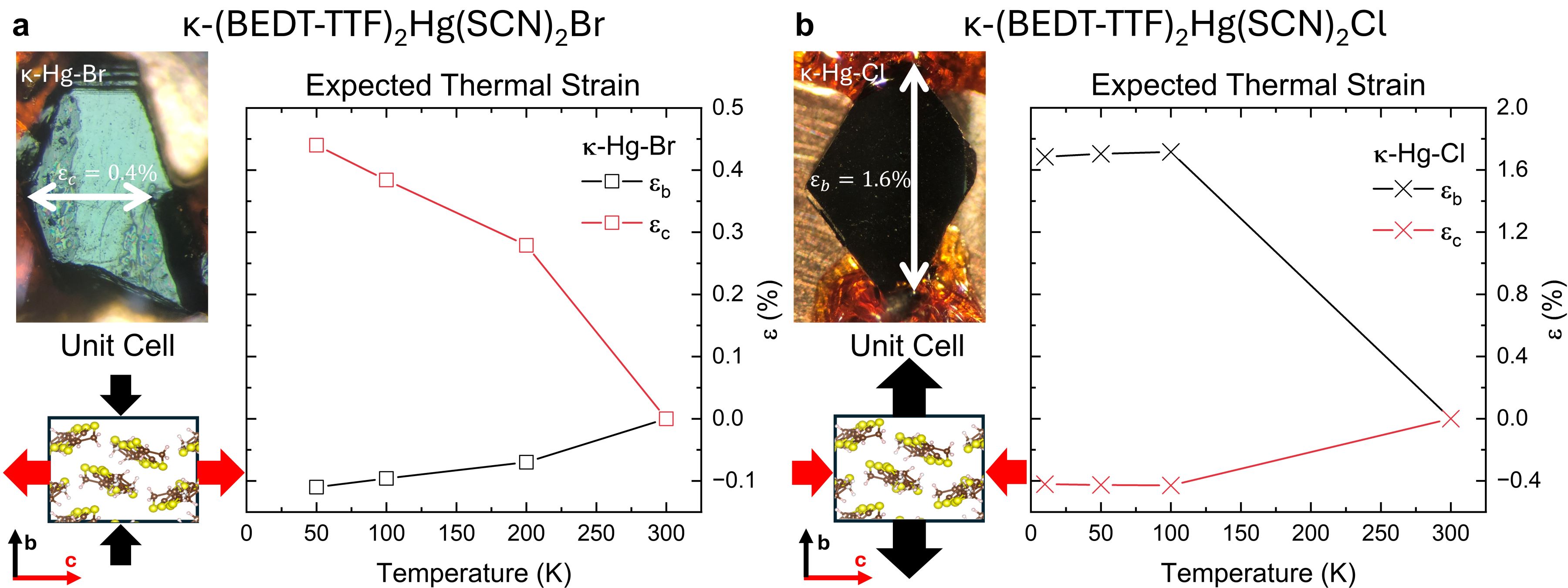}
    \caption{Scheme for Application of Strain
    {\bf (a)} $\kappa$-(BEDT-TTF)$_2$Hg(SCN)$_2$Br - the sample is glued along the c-axis on a copper substrate. The temperature dependence of expected thermal strain in $\kappa$-(BEDT-TTF)$_2$Hg(SCN)$_2$Br along the c-axis from gluing and b-axis from an estimated Poisson ratio of 0.25. {\bf (a)} $\kappa$-(BEDT-TTF)$_2$Hg(SCN)$_2$Cl - the sample is glued along the b-axis on a copper substrate. The temperature dependence of expected thermal strain in $\kappa$-(BEDT-TTF)$_2$Hg(SCN)$_2$Br along the b-axis from gluing and c-axis from an estimated Poisson ratio of 0.25}
    \label{Samples}
\end{figure*}

Samples of $\kappa$-(BEDT-TTF)$_2$Hg(SCN)$_2$X (X=Br, Cl) are small and fragile. Single crystal samples of these materials are typically plate-like with in-plane side lengths around 1~mm in length. Upon cooling, these materials exhibit large thermal contraction which greatly exceeds that of most inorganic materials. 
We applied uniaxial strain by using the difference in thermal contraction between the material, (see Tables~\ref{Br-UnitCell}~and~\ref{Cl-UnitCell}) and the Cu substrate~\cite{Rubin1954}.
We glued opposing ends of the crystal along each axis using GE varnish, as shown in Fig.~\ref{Samples}.
The applied strain is temperature dependent. As the Cu substrate cools to a temperature $T$, the change in the distance between the two glued ends $L_0$ is determined by the thermal contraction of Cu on cooling, $C_{\text{Cu}}(T)$.
\begin{equation}
L_0(T) = L_0(1+C_{\text{Cu}}(T))
\end{equation}
As a sample of $\kappa$-Hg-X cools to a temperature $T$ without strain, the change in the size of the sample along an axis $i$, $L_i^{(0)}$ is determined by the thermal contraction of the material, $C^{(0)}_{\text{X}, \  i}(T)$.
\begin{equation}
a_i^{(0)}(T) = L_i^{(0)}(1+C^{(0)}_{\text{X}, \ i}(T))
\end{equation}
The applied strain on the sample of a material X, along an axis $i$, at each temperature, $\epsilon_{\text{X}, \ i}(T)$, can be determined as follows:
\begin{equation}
\epsilon_{\text{X}, \ i}(T) = \frac{\Delta L_i}{L_i^{(0)}} = C_{\text{Cu}}(T) - C^{(0)}_{\text{X}, \ i}(T)
\end{equation}
where $\Delta L_i = L_0(T)-L_i^{(0)}(T)$ is the applied displacement along the axis $i$.
Since Cu contracts less on cooling than samples of $\kappa$-Hg-X, $C_{\text{Cu}}(T) - C^{(0)}_{\text{X}, \ i}(T) > 0 $ and we apply tensile strain. For X=Br, we apply a strain of 0.4\% along the c-axis. For X=Cl, we apply a strain of 1.6\% along the b-axis. While the strain is temperature dependent, we expect the strain to be roughly uniform below 100~K. 

\section{Crystallographic Information }

\begin{table}[h]
    \centering
    \begin{tabular}{c|c|c|c|c|c}
         Temperature (K) & a (\AA) & b (\AA) & c (\AA) & $\beta$ ($^\circ$) & V\\
         \hline 100	& 36.7093 & 8.2015 & 11.6494 & 90.16 & 3507.3\\
                200	& 36.8904 & 8.2619 & 11.6734 & 90.228 & 3557.9\\
                300	& 37.0923 & 8.3278 & 11.7296 & 90.345 & 3623.2
    \end{tabular}
    \caption{Temperature dependence of the unit cell of $\kappa$-(BEDT-TTF)Hg(SCN)$_2$Br}
    \label{Br-UnitCell}
\end{table}
\begin{table}[h]
    \centering
    \begin{tabular}{c|c|c|c|c|c}
         Temperature (K) & a (\AA) & b (\AA) & c (\AA) & $\beta$ ($^\circ$) & V\\
         \hline 10 & 36.367 & 8.1209 & 11.741 & 90.186 & 3467.5\\
                50 & 36.403 & 8.1226 & 11.7658 & 90.268 & 3478.9\\
                100 & 36.2617 & 8.1222 & 11.7248 & 90.266 & 3453.2\\
                300 & 36.5964 & 8.2887 & 11.7503 & 90.067 & 3564.2\\
    \end{tabular}
    \caption{Temperature dependence of the unit cell of $\kappa$-(BEDT-TTF)Hg(SCN)$_2$Cl~\cite{Drichko2014}}
    \label{Cl-UnitCell}
\end{table}
Both $\kappa$-(BEDT-TTF)$_2$Hg(SCN)$_2$Br ($\kappa$-Hg-Br) and $\kappa$-(BEDT-TTF)$_2$Hg(SCN)$_2$Cl ($\kappa$-Hg-Cl) crystallize in the C2/c space group (No. 15). The temperature dependence of the unit cell parameters of $\kappa$-Hg-Br and $\kappa$-Hg-Cl is shown in Table~\ref{Br-UnitCell} and Table~\ref{Cl-UnitCell}, respectively. The structural refinement of $\kappa$-Hg-Br at 100~K is given in Table~\ref{Br-Str-Ref}.

\begin{table}[h]
    \centering
    \begin{filecontents*}{SM/Br-100K.csv}
Atom,x,y,z,Occ.,U,Wyck. Pos.
Br(21),0,0.7951,0.25,1,0.027,4e
C(1),0.2656,0.5906,-0.08862,1,0.017,8f
C(2),0.30105,0.5852,-0.05344,1,0.017,8f
C(3),0.19721,0.5349,-0.12135,1,0.017,8f
C(4),0.20716,0.6386,-0.20622,1,0.017,8f
C(5),0.36303,0.4919,0.0279,1,0.018,8f
C(6),0.37134,0.5977,-0.0571,1,0.02,8f
C(7A),0.12742,0.5345,-0.2166,0.658,0.02,8f
C(7B),0.12587,0.581,-0.166,0.342,0.02,8f
C(8A),0.13729,0.7057,-0.2555,0.658,0.02,8f
C(8B),0.13579,0.6113,-0.2906,0.342,0.02,8f
C(9A),0.42869,0.5505,0.1219,0.789,0.02,8f
C(9B),0.4328,0.5671,0.102,0.211,0.02,8f
C(10A),0.44636,0.5888,-8E-4,0.789,0.02,8f
C(10B),0.444,0.5825,0.0262,0.211,0.02,8f
C(21),0.04443,0.4333,0.01737,1,0.024,8f
H(7AA),0.1302,0.4599,-0.2829,0.658,0.024,8f
H(7AB),0.1014,0.5338,-0.1948,0.658,0.024,8f
H(7BA),0.1008,0.5362,-0.1631,0.342,0.024,8f
H(7BB),0.1257,0.6869,-0.1251,0.342,0.024,8f
H(8AA),0.1362,0.78,-0.1887,0.658,0.024,8f
H(8AB),0.119,0.7432,-0.3123,0.658,0.024,8f
H(8BA),0.1157,0.6727,-0.3276,0.342,0.024,8f
H(8BB),0.1378,0.5049,-0.3302,0.342,0.024,8f
H(9AA),0.4173,0.651,0.1523,0.789,0.024,8f
H(9AB),0.4482,0.5173,0.176,0.789,0.024,8f
H(9BA),0.4226,0.6734,0.1265,0.211,0.024,8f
H(9BB),0.4523,0.5386,0.1571,0.211,0.024,8f
H(10A),0.4592,0.49,-0.0281,0.789,0.024,8f
H(10B),0.4648,0.676,0.0087,0.789,0.024,8f
H(10C),0.4521,0.4746,-0.002,0.211,0.024,8f
H(10D),0.4661,0.6504,0.0365,0.211,0.024,8f
Hg(21),0,0.48924,0.25,1,0.022,4e
N(21),0.04639,0.5111,-0.06526,1,0.031,8f
S(1),0.2309,0.48233,-0.02106,1,0.018,8f
S(2),0.25181,0.71141,-0.20397,1,0.02,8f
S(3),0.31653,0.46982,0.06224,1,0.017,8f
S(4),0.33466,0.69522,-0.1247,1,0.022,8f
S(5),0.1542,0.45175,-0.09519,1,0.026,8f
S(6A),0.18144,0.7183,-0.3181,0.658,0.018,8f
S(6B),0.178,0.7234,-0.3115,0.342,0.032,8f
S(7A),0.39503,0.39204,0.11446,0.789,0.019,8f
S(7B),0.3935,0.4007,0.117,0.211,0.058,8f
S(8A),0.41459,0.6505,-0.1079,0.789,0.024,8f
S(8B),0.412,0.689,-0.1,0.211,0.032,8f
S(21),0.04236,0.31689,0.13469,1,0.027,8f
\end{filecontents*}
\csvautotabular{SM/Br-100K.csv}
    \caption{Structural Refinement of $\kappa$-(BEDT-TTF)$_2$Hg(SCN)$_2$Br at 100~K. The numeric label associated with each atom denotes the corresponding carbon site which the atom is bonded to. The alphabetical label denotes the position of an atom in a given conformation. The relative abundance of each conformation is denoted by the occupancy.}
    \label{Br-Str-Ref}
\end{table}
\newpage

\section{Vibrational spectra analysis }
\subsection{Fitting of Molecular Vibrations and Lattice Phonons}
We fit Raman active molecular vibrations and lattice phonons using Voigt profiles, or a convolution of a Gaussian distribution with a Lorentzian profile: 
\begin{equation}
    I(\omega) \propto \int_{-\infty}^{\infty}d\omega'G(\omega', \omega, \sigma_\text{res})L(\omega, \omega_i, \Gamma_i)
\end{equation}
The $\sigma_\text{res}$ parameter is the Gaussian spectral resolution parameter. For all measurements, this value is 2~\cm. $\omega_i$ is the frequency of vibration. $\Gamma_i$ is the phenomenological damping factor, which is inversely proportional to the lifetime of a vibration~\cite{Klemens1966}. 
We fit each molecular vibration and lattice phonon individually and then fit all of the lineshape parameters over a frequency range of 200~\cm ~together with a linear background to approximate broad features, such as photoluminescence and soft mode of dipole fluctuations~\cite{Hassan2018, Hassan2018Crystals}.

\subsection{Analysis of Charge Sensitive Molecular Vibrations}

To identify the electronic ground state, we use Raman Scattering Spectroscopy to measure the Raman active charge sensitive molecular vibration $\nu_2$.
For this charge sensitive molecular vibration, we analyze both the frequency and linewidth. The measured frequency corresponds to charge of the BEDT-TTF molecule, denoted by $n$, as follows~\cite{Yamamoto2005}:
\begin{equation}
    \omega(n) = 1447\text{\cm} + 120 \text{\cm}(1-n)
    \label{freq-Eq}
\end{equation}
 If there are two spectral profiles for $\nu_2$ with a difference in frequency of $\Delta\omega$, we infer that there are charge rich and charge poor BEDT-TTTF molecules. We assign the charge on each BEDT-TTF molecule (in units of electron charge) as follows:
\begin{equation}
    n_{\pm}(\Delta\omega) \ = \ 0.5  \ \pm \   \frac{\Delta\omega}{240 \text{\cm}}
    \label{charge-Eq}
\end{equation}


{\bf Static Charge Order: }
In the case of static charge order, our analysis follows the splitting of the spectral lineshape of $\nu_2$ from one profile into two distinct profiles. At temperatures above the charge order transition, we fit a single broad Voigt profile. In the charge ordered state, we fit two independent narrow Voigt profiles, with a difference in frequency (parameterized by $\Delta\omega$) determined by the charge separation $\Delta n=n_+(\Delta\omega)-n_-(\Delta\omega)$  (See Eq.~\ref{charge-Eq})~\cite{Yamamoto2005}.\\

{\bf Fluctuating Charge Order: }
In the case of dynamic charge order, charge fluctuates between charge rich and charge poor BEDT-TTF molecules. In this case we always use a nonzero charge separation in our analysis regardless of whether we observe the splitting of the spectral lineshape of $\nu_2$.
We use the phenomenological two-site model with fluctuations to simulate the lineshape~\cite{Kubo1969, Yakushi2012_Crystals}:
\begin{widetext}
{\footnotesize 
\begin{equation}
    I(\omega) \propto \text{Re} \left [ \begin{pmatrix} a_1 & a_2
    \end{pmatrix} 
    \begin{pmatrix}
        \mathrm{i}(\omega - (\omega_0 - \frac{\Delta\omega}{2}
        )) + \frac{\omega_{\text{ex}}}{2} + \Gamma_0/2 & -\frac{\omega_{\text{ex}}}{2}
        \\
        -\frac{\omega_{\text{ex}}}{2} & \mathrm{i}(\omega - (\omega_0 + \frac{\Delta\omega}{2}
        )) + \frac{\omega_{\text{ex}}}{2} + \Gamma_0/2
    \end{pmatrix}^{-1}
    \begin{pmatrix}
        a_1 \\
        a_2
    \end{pmatrix}
    \right ]
\end{equation}}
\end{widetext}
Where $\mathrm{i}$ is the imaginary unit. $a_1$ and $a_2$ are intensity prefactors for sites 1 and 2, respectively.
$\omega_0$ is the frequency of $\nu_2$ without charge disproportionation.
$\Gamma_0$ is the lifetime of the pure vibration,  which is fixed using the linewidth of $\nu_3$. 
$\Delta\omega$ is the splitting in frequency of the two spectral profiles. $\omega_{\text{ex}}$ is the frequency of fluctuations of charge between charge rich and charge poor molecules. The charge on each molecule is calculated using Eq.~\ref{charge-Eq}. \\

The measured spectral linewidth of the charge sensitive mode $\nu_2$ depends on the lifetime of the uncoupled molecular vibration and charge fluctuations.
In the absence of charge fluctuations with only phonon-phonon coupling, the lifetime of the molecular vibration $\nu_2$ follows the Klemens' law of anharmonic decay~\cite{Klemens1966}.
In this case, the lifetime is experimentally determined by the linewidth of the narrow spectral lineshape of $\nu_3$, which is not coupled to the charge on the BEDT-TTF molecule.
Without charge fluctuations, as temperature decreases, the lifetime of the vibration increases, leading to a narrowing of the spectral lineshape, as observed for the $\nu_3$ mode. 
Therefore, an increase of the linewidth of $\nu_2$ at low temperatures is interpreted as the result of charge fluctuations~\cite{Yamamoto2005, Yakushi2012_Crystals, Hassan2018, Liebman2024}.
These charge fluctuations correspond to charges hopping between charge rich and charge poor BEDT-TTF molecules within (BEDT-TTF)$_{2}^{1+}$ dimer lattice sites. As the amplitude of charge fluctuations increases (parameterized by $\Delta\omega$), the linewidth of $\nu_2$ increases. As the lifetime of charge fluctuations increases (parameterized by $1/\omega_{\text{ex}}$), the linewidth of $\nu_2$ also increases. Therefore, the broadening of the lineshape of $\nu_2$ is an indirect measurement of charge fluctuations.

\section{Treatment of Spectra}

\begin{figure}[h]
    \centering
    \includegraphics[width=\linewidth]{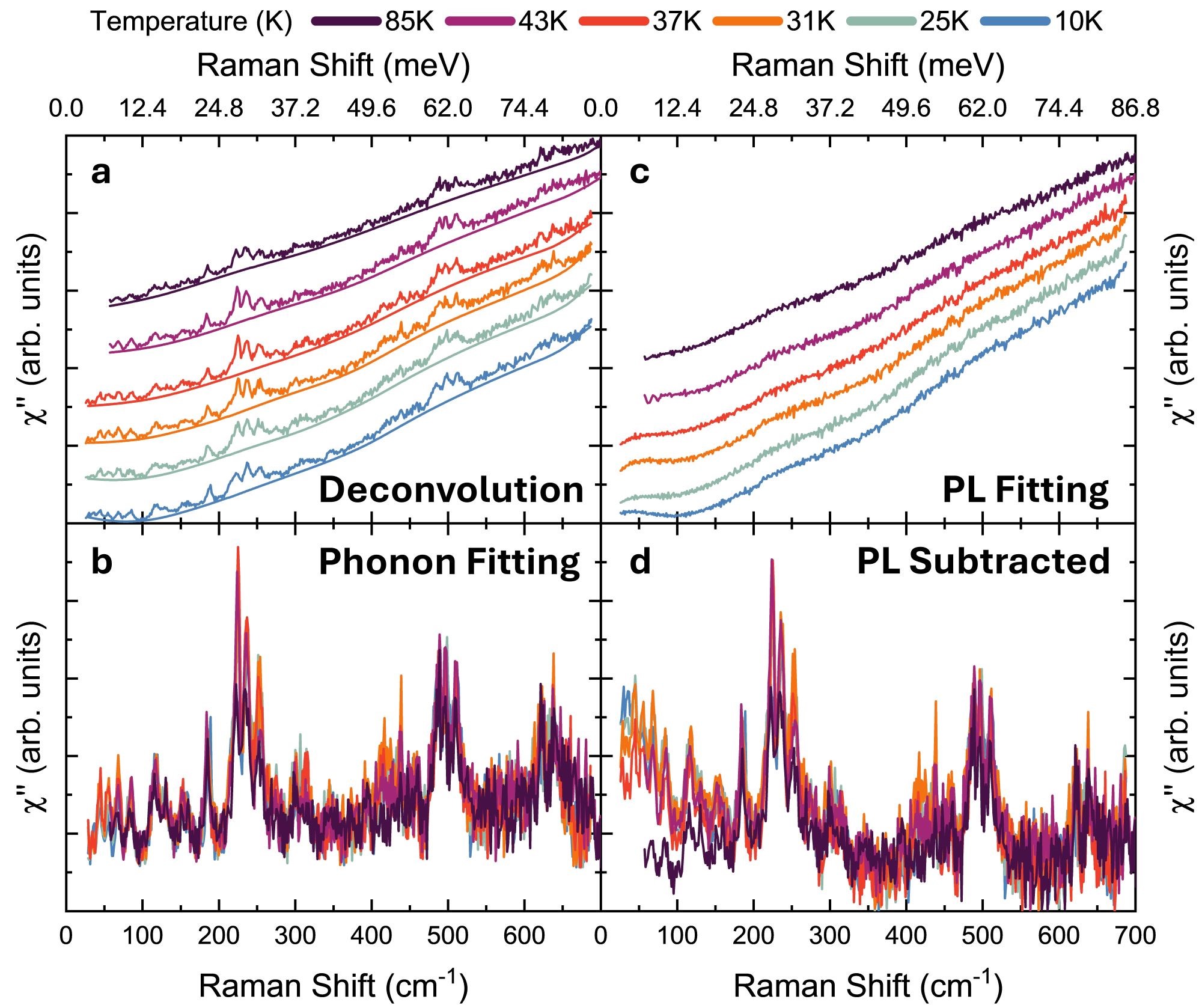}
    \caption{Treatment of spectra of $\kappa$-(BEDT-TTF)$_2$Hg(SCN)$_2$Br. {\bf a} Raw spectra, with estimated background for background isolation. {\bf b} Peak spectra used for Phonon Fitting. {\bf c} Phonon Subtracted spectra used for PL fitting in the frequency range above 200\cm. {\bf d} Spectra with PL subtracted.}
    \label{Br-Spec-Treatment}
\end{figure}

\begin{figure}[h]
    \centering
    \includegraphics[width=\linewidth]{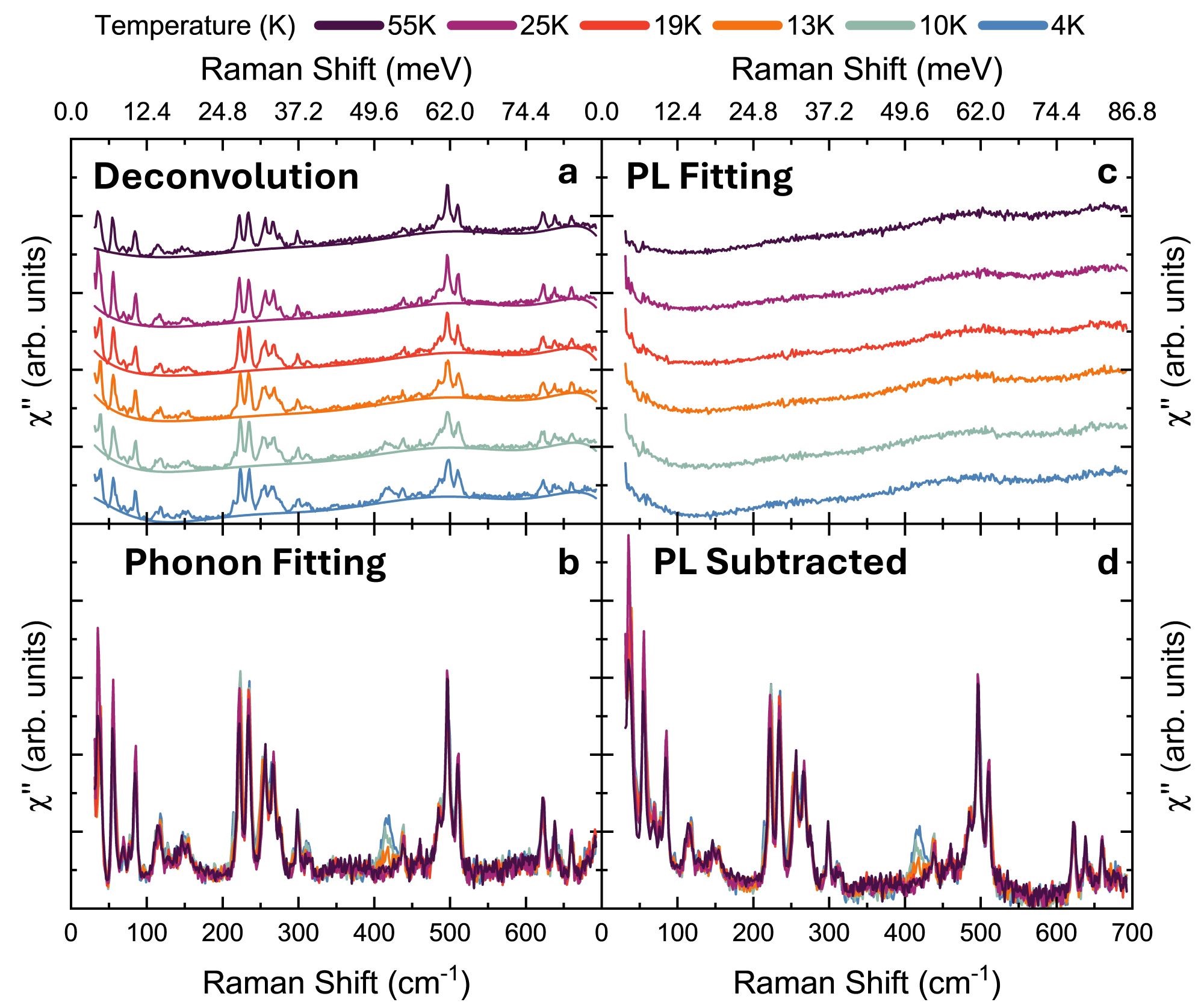}
    \caption{Treatment of spectra of $\kappa$-(BEDT-TTF)$_2$Hg(SCN)$_2$Cl. {\bf a} Raw spectra, with estimated background for background isolation. {\bf b} Peak spectra used for Phonon Fitting. {\bf c} Phonon Subtracted spectra used for PL fitting in the frequency range above 200\cm. {\bf d} Spectra with PL subtracted.}
    \label{Cl-Spec-Treatment}
\end{figure}

The spectra are treated to remove photoluminescence (PL). PL occurs following the absorption of light, when an electron or hole is excited to a real state, and emits a photon corresponding to the transition to the ground state~\cite{Mooradian1969, Shinde2012}. This process is different from Raman scattering, where electrons are excited to virtual states. 
In BEDT-TTF molecules and BEDT-TTF based materials~\cite{Kozlov1995}, PL has been measured in the range 1.6-2~eV, corresponding to red and infrared wavelengths, appearing as a double peak structure with a Lorentzian lineshape. 
In our measurement, we excite the material with green light (514.5 nm or 2.41 eV), which can excite PL with longer wavelengths or less energy, which includes the known energy range of PL~\cite{Kozlov1995}. 
Since PL occurs at wavelengths that are red to infrared, the PL spectra are centered at frequencies greater than the upper limit of the spectral window of our measurements. In our present work, we fit the PL to a broad Lorentzian centered above 700~cm$^{-1}$, the upper limit of the measured spectral window (See Figs.~\ref{Br-Spec-Treatment},\ref{Cl-Spec-Treatment}). 

\begin{figure}[h]
    \centering
    \includegraphics[width=0.75\linewidth]{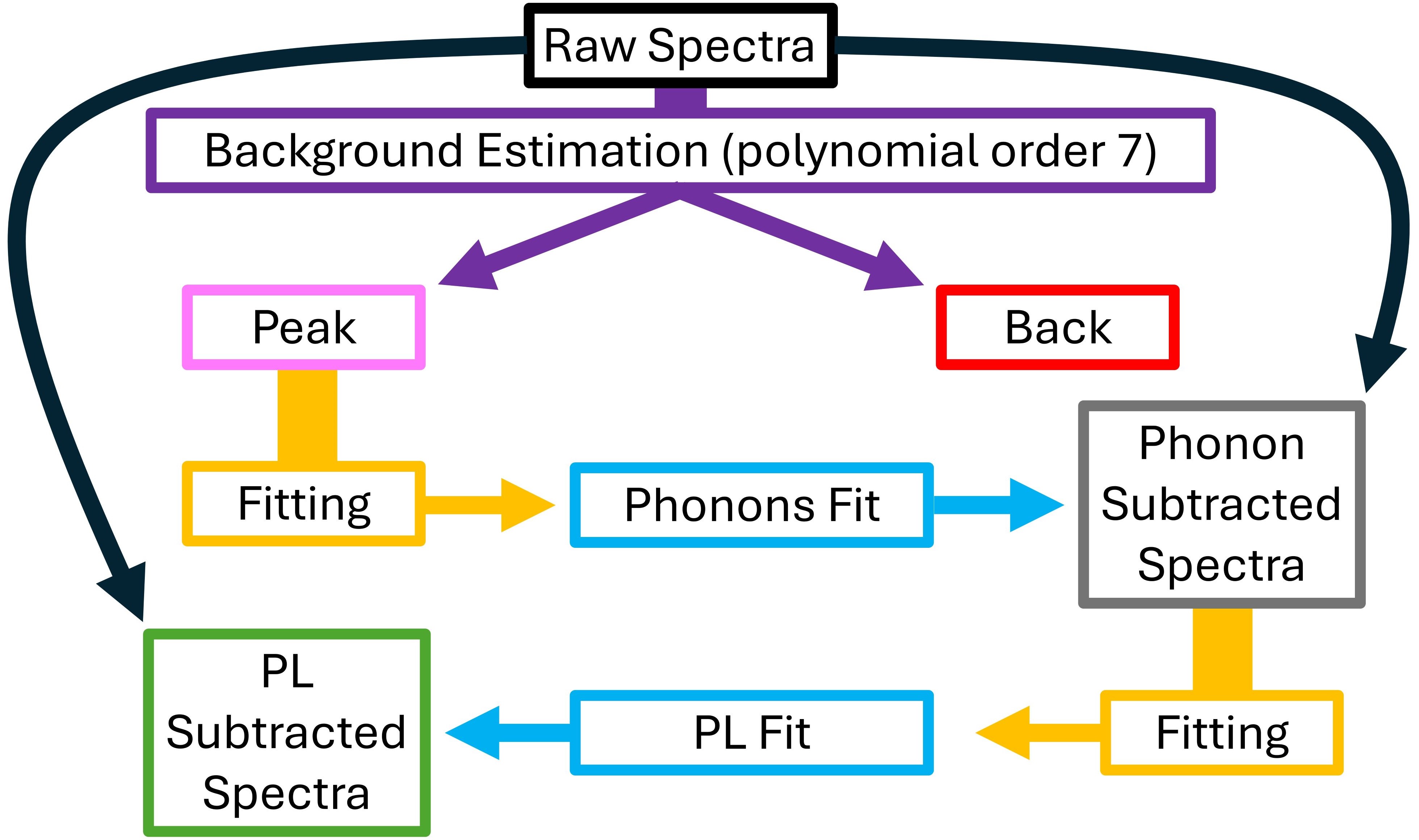}
    \caption{Scheme for treatment of spectra using the iterative modified polynomial fitting, denoting each step taken from the raw spectra to the final PL subtracted spectra.}
    \label{Spec-Treatment}
\end{figure}

We isolate broad background features and narrow vibrations using an iterative modified polynomial fitting procedure~\cite{Zhao2007}. 
The scheme outlying this process is shown in Fig.~\ref{Spec-Treatment}.
The background estimation was conducted with a polynomial of degree 7 and a convergence parameter of $10^{-4}$. 
We show each step in the process, from estimation of the background from the raw data to the final PL subtracted data.
For $\kappa$-Hg-Br this process is shown with the corresponding spectra in Fig.~\ref{Br-Spec-Treatment}. For $\kappa$-Hg-Cl this process is shown with the corresponding spectra in Fig.~\ref{Cl-Spec-Treatment}. Starting from the raw spectra (Fig.~\ref{Br-Spec-Treatment}{\bf a} and Fig.~\ref{Cl-Spec-Treatment}{\bf a}), we isolate all background features from vibrations~\cite{Zhao2007}. There are individual spectra for backgrounds (Back), which are plotted underneath the raw data in Fig.~\ref{Br-Spec-Treatment}{\bf a} and Fig.~\ref{Cl-Spec-Treatment}{\bf a}, and spectra for vibrations (Peak), which we plot in Fig.~\ref{Br-Spec-Treatment}{\bf b} and Fig.~\ref{Cl-Spec-Treatment}{\bf b}. The vibrational spectra are fit by a sum of Voigt profiles with a flat background, consistent with the analysis procedure of vibrational bands described in Section~{\bf IV. Vibrational Spectra Analysis}.
The frequency of each mode corresponds to a known vibrational frequency from IR and Raman measurements~\cite{Drichko2014, Ivek2017, Hassan2018, Hassan2020, Liebman2024}. The fitted phonon spectra are then subtracted from the Raw Spectra, which results in the Phonon Subtracted Spectra, which we show in Fig.~\ref{Br-Spec-Treatment}{\bf c} and Fig.~\ref{Cl-Spec-Treatment}{\bf c}.

We use the Phonon Subtracted Spectra to distinguish broad features, including collective dipole fluctuations and PL.
As shown in Fig.~\ref{Br-Spec-Treatment}{\bf c}, the collective dipole fluctuations of $\kappa$-Hg-Br appear as a broad feature below 200~cm$^{-1}$. The PL corresponds to a broad background which extends past the highest frequencies in the spectral window.
For $\kappa$-Hg-Cl (Fig.~\ref{Cl-Spec-Treatment}{\bf c}), the low frequency range contains a large contribution from the elastic line, which prevents any possible identification of broad low frequency excitations. However, there is PL, which corresponds to the broad background that extends past the largest measured frequencies. For both materials $\kappa$-Hg-Br and $\kappa$-Hg-Cl, the PL is then fit to a Lorentzian lineshape, using the high frequency region of the Phonon Subtracted Spectra. Then the fit PL lineshape is subtracted from the Raw Spectra. Finally, we have the PL subtracted spectra, where collective dipole fluctuations and vibrations can be examined independently of PL. 
The the PL subtracted spectra are shown for $\kappa$-Hg-Br in Fig.~\ref{Br-Spec-Treatment}{\bf d} and for $\kappa$-Hg-Cl in Fig.~\ref{Cl-Spec-Treatment}{\bf d}.

\newpage

\newpage
\newpage
\section{Low Frequency Spectra of $\kappa\text{-(BEDT-TTF)}_2\text{Hg(SCN)}_2\text{Cl}$}
\begin{figure}[h]
    \centering
    \includegraphics[width=0.75\linewidth]{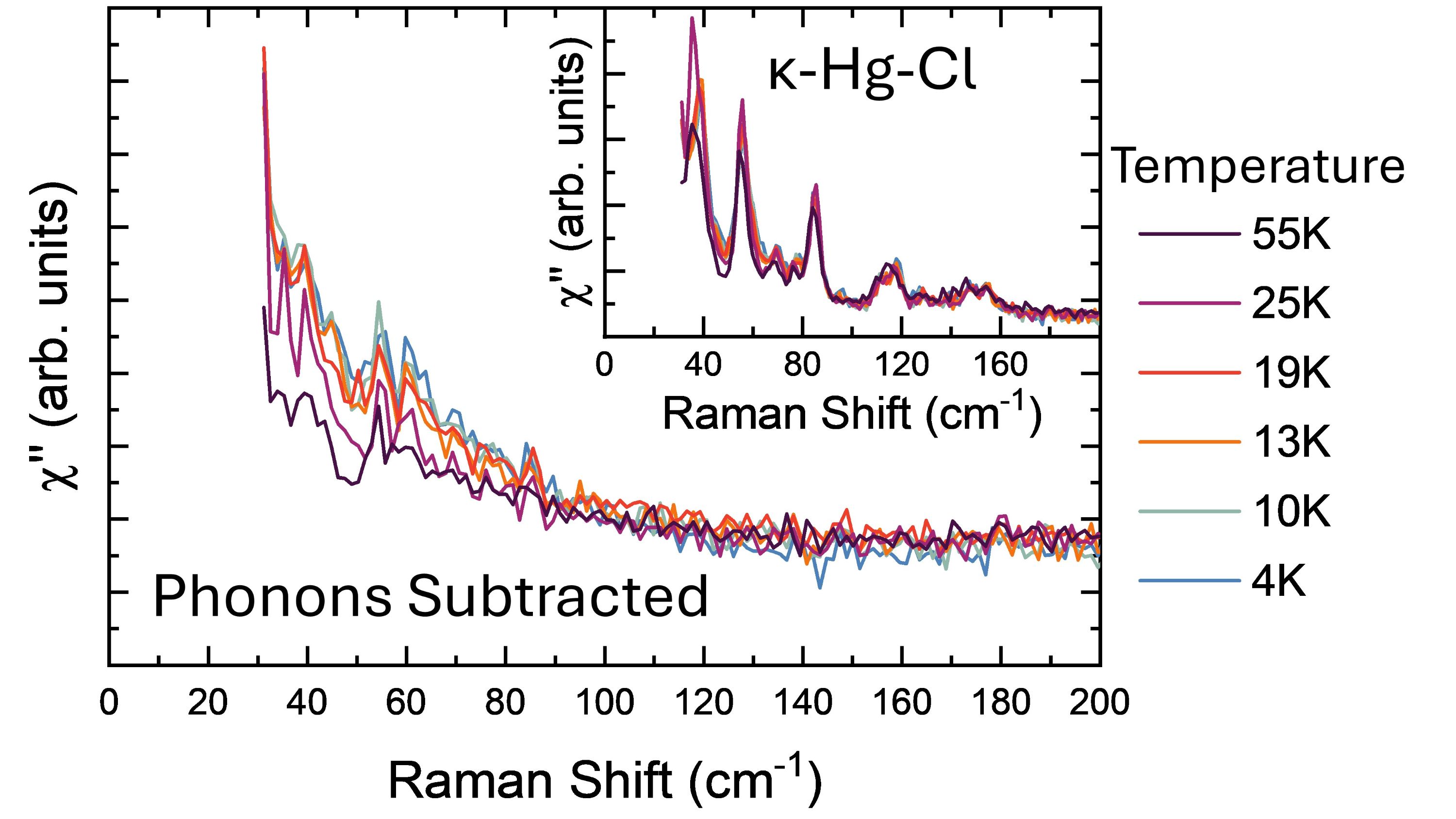}
    \caption{Low frequency Raman scattering spectra of $\kappa$-(BEDT-TTF)$_2$Hg(SCN)$_2$Cl in the (c,c) polarization, with phonons subtracted. Spectra without phonons subtracted are shown in the insets.}
    \label{Cl_DF}
\end{figure}

In the absence of strain, $\kappa$-Hg-Cl undergoes a metal to charge order insulator transition at T$_{\text{CO}}$=30~K~\cite{Drichko2014}, where collective dipole fluctuations are not observed~\cite{Hassan2018Crystals}.
We plot the low frequency Raman spectra of $\kappa$-Hg-Cl under strain, where T$_{\text{CO}}$=10~K, with phonons subtracted in the (c,c) polarization in Fig.~\ref{Cl_DF} (for the spectra without phonons subtracted see inset). 
While the presence of the elastic line precludes us from going to low frequencies, the changes in the observed range are negligible.


\section{Fitting Low Frequency Spectra of $\kappa\text{-(BEDT-TTF)}_2\text{Hg(SCN)}_2\text{Br}$}
The low frequency spectra of $\kappa\text{-(BEDT-TTF)}_2\text{Hg(SCN)}_2\text{Br}$ shows a broad mode below 200~\cm (see Fig.~\ref{Br-Spec-Treatment}). 
We make only 3 assumptions to the broad mode, which define a Lorentzian profile:
(1) The mode has a central frequency, $\nu$, or a frequency with the maximal intensity in Raman scattering. (2) The mode corresponds to excitations that have an average lifetime, $\tau$, or an average decay rate of $\gamma=\tau^{-1}$. (3) The mode has a well-defined spectral weight at each temperature, $\text{I}(\text{T})$, which is equal to the integrated intensity of the mode in the Raman scattering spectra. 
We analyze this mode as a Lorentzian profile and interpret the feature as a collective excitation predicted for charge fluctuations within (BEDT-TTF)$_2^{1+}$ dimers, or a collective dipole fluctuation~\cite{Hotta2010, Naka2010, Yao2018}. The mode is broad, and therefore not resolution limited; however, to remain consistent with all fitting procedures, we consider the spectral resolution of 2\cm in the fitting procedure. We therefore fit the mode to a Voigt profile, where the fit parameters $\nu$, $\tau$, and $\text{I}(\text{T})$ correspond to the Lorenztian profile. Since a Voigt profile is normalized to $1$, the fit parameter for intensity, $\text{I}(\text{T})$, corresponds to the spectral weight of the Lorentzian profile.

\begin{figure}
    \centering
    \includegraphics[width=\linewidth]{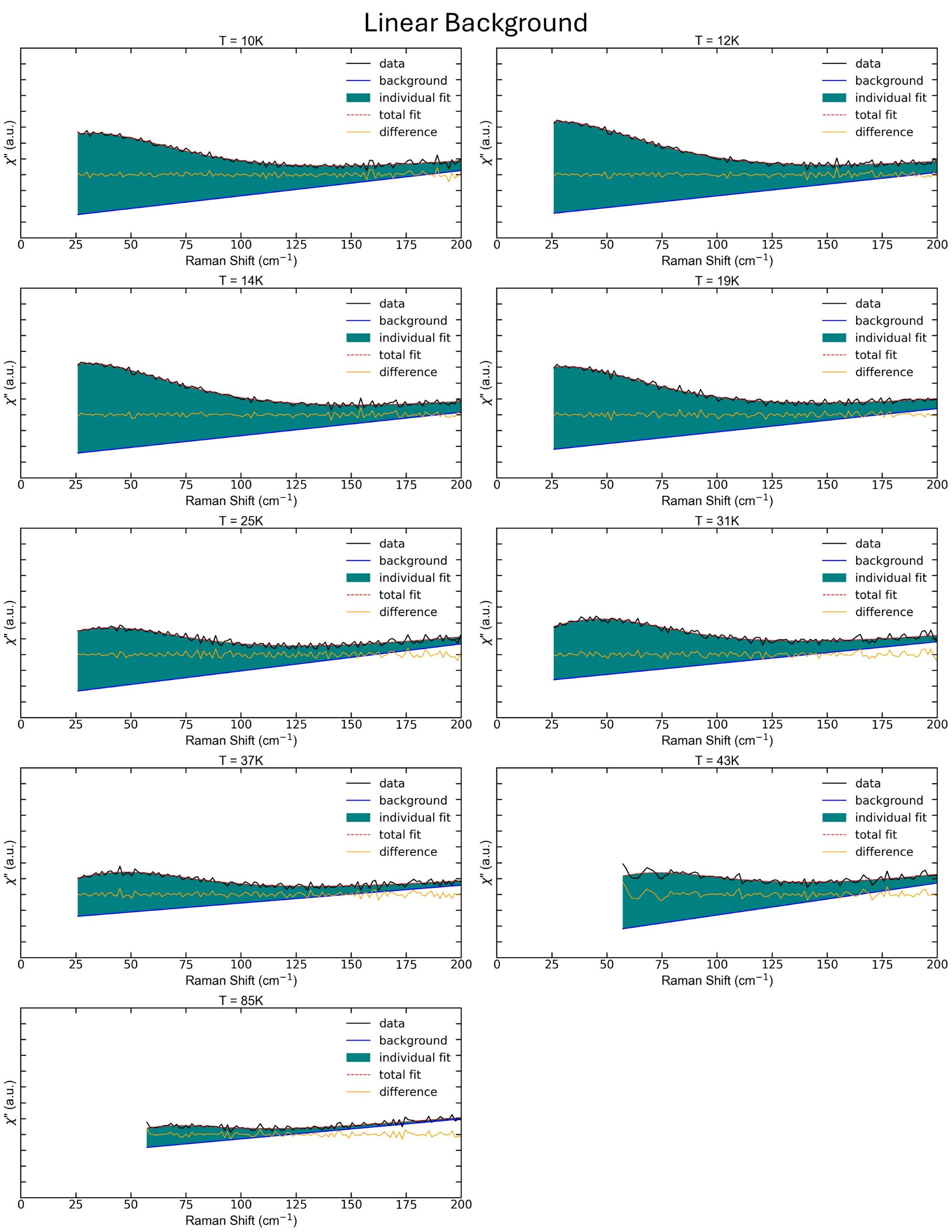}
    \caption{Fits of the low frequency Raman spectra of $\kappa\text{-(BEDT-TTF)}_2\text{Hg(SCN)}_2\text{Br}$ using a linear background}
    \label{Linear}
\end{figure}

\begin{figure}
    \centering
    \includegraphics[width=\linewidth]{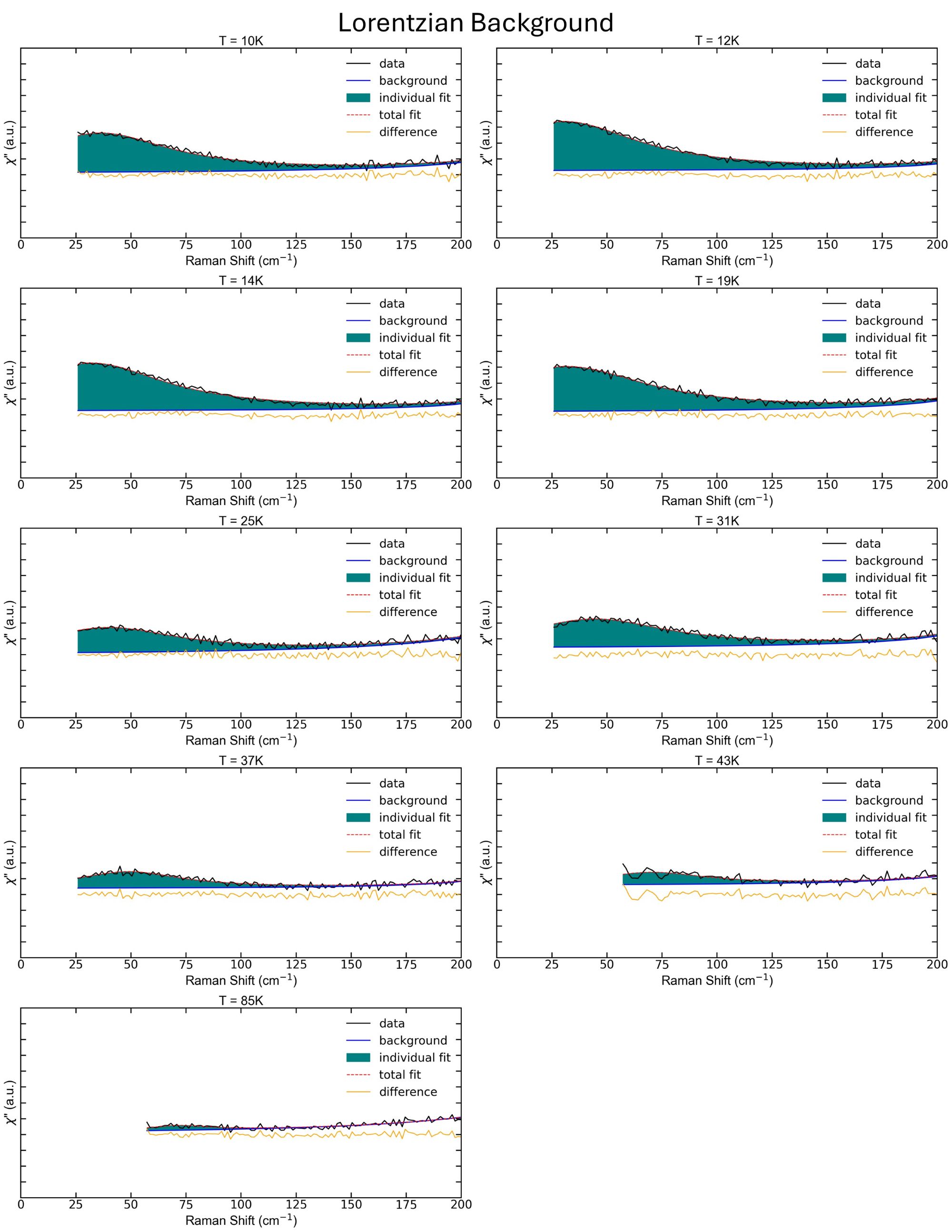}
    \caption{Fits of the low frequency Raman spectra of $\kappa\text{-(BEDT-TTF)}_2\text{Hg(SCN)}_2\text{Br}$ using a Lorentzian background}
    \label{Lorentzian}
\end{figure}

When fitting this mode, we are susceptible to systematic errors. These errors arise from the procedure of subtracting phonons and the high-frequency PL background (see section~{\bf V. Treatment of Spectra}). The latter is an optical excitation which is described by a Lorentzian, and far away from the center of the peak, a Lorentzian can be approximated to a line. When fitting the broad mode, we compare two different fitting procedures: (I) The background underneath the broad excitation is linear and (II) The background underneath the broad excitation is Lorentzian.
We show the results of the Linear background fitting procedure in Fig.~\ref{Linear} and the Lorentzian background fitting procedure in Fig.~\ref{Lorentzian}. For each fit, we plot the spectral weight of the fit of the low frequency mode with a teal shaded area.
In the first case, the fit of the low frequency mode exhibits the largest reasonable spectral weight at the lowest measured frequencies, as shown in Fig.~\ref{Linear}. 
Therefore the linear background results in the lower bound of the fitted central frequency.
We find that the Lorentzian background fits the data best. We use the discrepancies between the fitting procedures to estimate the systematic errors involved in the fitting procedure.

\newpage

\section{Tuning of Hubbard Parameters under Strain}

We now discuss the calculated tuning of the Extended Hubbard Model parameters under strain.
As shown in Fig.~\ref{Samples}, we calculate the expected strain given the difference in thermal contraction between the sample and the Cu substrate~\cite{Rubin1954}. We also show the calculated lattice constants under strain in Fig.~\ref{Samples}. We consider the Poisson Effect, which produces smaller strains along the other axes of the crystal. We assume the Poisson ratio to be roughly 0.25, which serves as a realistic estimate without direct measurement~{\cite{Greaves2011}}.

\begin{figure}
    \centering
    \includegraphics[width=0.6\linewidth]{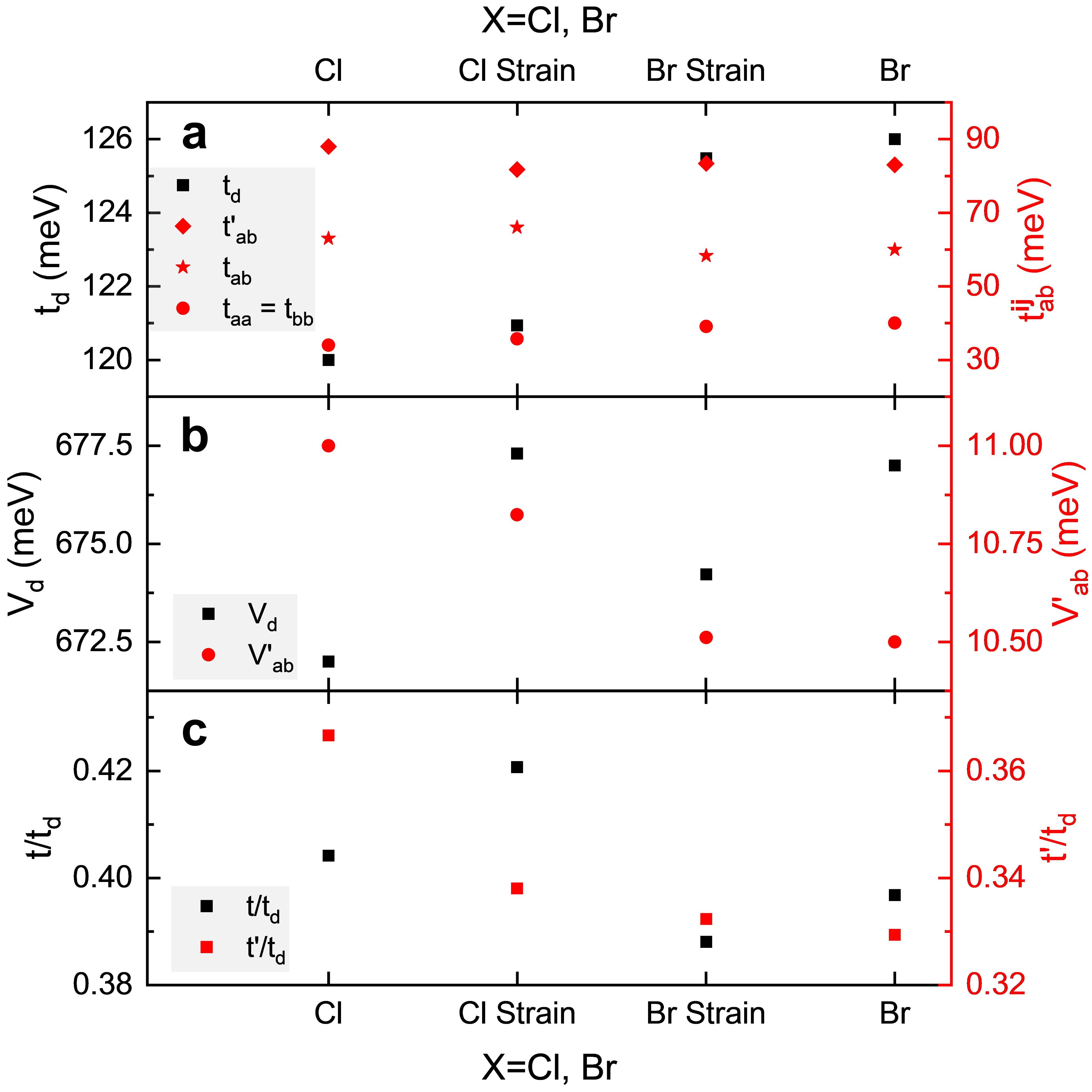}
    \caption{Calculated tuning of interactions via strain in $\kappa$-(BEDT-TTF)$_2$Hg(SCN)$_2$X (X=Br, Cl).
    At maximal and constant strain below 50~K: {\bf (a)} calculated intermolecular charge transfer integrals including intradimer charge transfer $t_d$ (left axis) and interdimer monomer charge transfer $t'_{ab}, \ t_{ab}, \ t_{aa} = t_{bb}$ {\bf (b)} intradimer electron repulsion and interdimer electron repulsion $V'_{ab}$ {\bf (c)} interdimer charge transfer integrals within the dimer approximation $t, \ t'$ in units of $t_d$.}
    \label{Str-Hub-Lat}
\end{figure}

We use the calculated strain to estimate the leading order effects to be tuning of the parameters of the Extended Hubbard Model. 
In particular we predict the shift to intersite electron repulsion and intersite hopping, parameterized through $V_{ab}^{ij}$ and $t_{ab}^{ij}$, respectively. We present the calculated Hubbard model parameters under strain in Fig.~\ref{Str-Hub-Lat}. For nearest neighboring interactions $t, V$, the interactions between sites $a$ and $b$ of different dimer sites are $V_{ab}=V_{ba}, \ t_{ab}=t_{ba}$, and between sites $a$ and $a$ or sites $b$ and $b$ are $V_{aa}=V_{bb}, \ t_{aa}=t_{bb}$. For the second nearest neighboring interactions $t', V'$, there is only interaction between inequivalent sites $V'_{ab}=V'_{ba}, \ t'_{ab}=t'_{ba}$. We use the following approximations for each Hubbard Model parameter, $x$, as a function of the applied strain $\epsilon$:
\begin{equation}\label{Eq. 9}
\begin{split}   
    x(\epsilon) = \ & \ \sum_{n=0}\epsilon^n\frac{d^nx}{d\epsilon^n}\rvert_{\epsilon=0} \\
    = \ & \  x(0) + \epsilon\frac{dx}{d\epsilon}\rvert_{\epsilon=0} + O(\epsilon^2)  \approx x_0 + \epsilon\frac{dx}{d\epsilon}\rvert_{\epsilon=0}
\end{split}
\end{equation}
For intersite repulsion, $V_{ab}^{ij}$, we expect the exact approximation of Eq.~\ref{Eq. 9} holds:
\begin{equation}
    V_{ab}^{ij}[\epsilon]=\frac{v_{ab}^{ij}}{r_{ab}^{ij}(1+\epsilon)} \approx \frac{v_{ab}^{ij}}{r_{ab}^{ij}}(1-\epsilon)=(1-\epsilon)(V_{ab}^{ij})[0]
    \label{Eq.-V-strain}
\end{equation}
For $t_{ab}^{ij}$, we make the following approximation:
\begin{equation}
\begin{split}
    t_{ab}^{ij}(\epsilon) \propto \ & \  \exp[-\frac{l_{ab}^{ij}(1+\epsilon)}{a_0}]\\
    \approx \ & \ \exp[-\frac{l_{ab}^{ij}}{a_0}](1-\epsilon\frac{l_{ab}^{ij}}{a_0})=(1-\epsilon\frac{l_{ab}^{ij}}{a_0})(t_{ab}^{ij})[0]
\end{split}
\label{Eq.-t-strain}
\end{equation}
where $l_{ab}^{ij}$ is the unstrained intermolecular distance between molecules $a,\ b$ in dimer sites $i,\ j$, and $a_0$ parameterizes the size of molecular orbitals in the tight binding approximation. For the approximated values in Fig.~\ref{Str-Hub-Lat}{\bf a} and Fig.~\ref{Str-Hub-Lat}{\bf c}, we let $\frac{l_{ab}^{ij}}{a_0}\approx 1$ when parameterizing $t_{ab}^{ij}=t_d$, the hopping between molecules in nearest-neighboring dimers.
We also calculate the renormalization of $ t_{ab}^{ij}= t_{ab}, \ t'_{ab}$, hopping between molecules from nearest and next-nearest neighboring dimers, respectively.
In $\kappa$-Hg-Br, the intradimer distance is 3.6~\AA~and the interdimer distances are 7.1~\AA and 8.3~\AA~\cite{Aldoshina1993}. The structural refinement of $\kappa$-Hg-Br at 100K is provided in Table~\ref{Br-Str-Ref}. In $\kappa$-Hg-Cl, the intradimer distance is 3.5~\AA~ and the interdimer distances are 7.1~\AA and 8.1~\AA\cite{Konovalikhin1992}. 
Therefore, we expect the fraction $\frac{l^{ij}_{ab}}{a_0}$, and subsequent strain dependence is roughly twice as large for $t_{ab}$ and $t'_{ab}$ when compared to $t_d$.
The ET molecules are oriented such that the long molecular axis and interatomic distances are mostly along the $a$-axis. We therefore expect a much smaller renormalization of $U$, so we allow all parameters other than the Hubbard $U$ to be strain dependent.
\section{Monte Carlo Simulation of Transverse Field Ising Model}
As discussed in the main text, emergent electric dipoles within (ET)$_2^{1+}$ are mapped onto an $S=1$ Transverse Field Ising Model (TFIM)~\cite{Naka2010, Hotta2010, Jacko2020}. We write the TFIM as follows:
\begin{equation}
    \hat{H} = K_\perp\sum_{i}\hat{S}_i^x+\frac{1}{2}\sum_{i\neq j}K^{ij}\hat{S}_i^z\hat{S}_j^z
\end{equation}
Following Ref.~\cite{Jacko2020}, we use the following parameterization:\\
\begin{align}
    K_\perp \ & = \ 2t_d\\
    K^{ij} \ & = \ D^{ij} + \tau^{ij}\\
    D^{ij} \ & = \ V^{ij}_{aa} + V^{ij}_{bb} - V^{ij}_{ab} - V^{ij}_{ba}\\
    \tau^{ij} \ & = \ \frac{(t^{ij}_{ab})^2-(t^{ij}_{aa})^2}{V_d}-8V_d\frac{(t^{ij}_{ab})^2-(t^{ij}_{aa})^2}{D^2}\\
    D \ & = \ \frac{77V_d^2}{100}
\end{align}
In the notation used above, $V^{ij}_{aa}$ refers to the Coulombic repulsion between molecular sites $a$ and $a$ between dimer lattice sites $i$ and $j$ and $V^{ij}_{ab}$ refers to the Coulombic repulsion between molecular sites $a$ and $b$ between dimer lattice sites $i$ and $j$. For hopping terms $t$, it is the same as $V$.
To perform a Monte Carlo simulation, we rewrite the local state 
\begin{equation}
    \ket{S_i} = \sin\frac{\theta}{2}\ket{+z}+\cos\frac{\theta}{2}\ket{-z}=\ket{\theta}
\end{equation}
This allows us to write a few observables:\\
\begin{align}
    \langle \hat{S}_x \rangle_\theta \ & = \ \sin\theta\\
    \langle \hat{S}_z \rangle_\theta \ & = \ \cos\theta\\
    \langle \hat{S}_x^2 \rangle_\theta \ & = \ \sin^2\theta\\
    \langle \hat{S}_z^2 \rangle_\theta \ & = \ \cos^2\theta
\end{align}
We also can rewrite the Hamiltonian in terms of the value of $\theta$ at each lattice site $i$:
\begin{equation}
    H = K_\perp\sum_i\sin\theta_i + \frac{1}{2}\sum_{i\neq j}K^{ij}\cos\theta_i\cos\theta_j
\end{equation}
This approach allows us to perform a classical Monte Carlo simulation, where we can evaluate the expectation values of  $\langle \hat{S}_x^2 \rangle$ and $ \langle \hat{S}_z^2 \rangle$, which can differentiate between the dimer Mott and charge order phases. We use a $16\times 16$ cluster with periodic boundary conditions and evaluate $ \langle \hat{S}_z^2 \rangle$ at T=0 to identify the ground state for a wide selection of parameters for $K_1$ and $K_2$ in the range corresponding to the calculated parameterization of $K_1$ and $K_2$ in Ref~\cite{Jacko2020} and the calculated tuning of interactions via strain, shown in Fig.~\ref{Str-Hub-Lat}. We use $4\times N^2 \times 1000 = 4\times 16^2\times 1000 = 1024000$ iterations to effectively sample an average 4000 values of $\theta$ for each of the 256 sites. At each iteration we use the following procedure:
\begin{enumerate}
    \item Randomly sample a lattice site
    \item Randomly choose a new value of $\theta \in [0, 2\pi)$
    \item Calculate the difference between the current and new energies, $\Delta E = E_{\text{new}}-E_{\text{old}}$
    \item If the new energy is lower ($\Delta E < 0$), accept the new value of $\theta$. If not, choose a random number from 0 to 1 and see if this number is less than $\exp(-\Delta E \beta )$. For $T=0$, an energy gain is always rejected.
\end{enumerate}
At the end of the simulation, the expectation values, such as $\langle \hat{S}_z^2 \rangle$ are calculated.
\newpage
\section{Bibliography}
\bibliography{SM.bbl}